\documentclass[11pt]{article}
\pdfoutput=1
\usepackage{jcapmod}

\usepackage{shorthand}
\usepackage{mathtools}
\usepackage{booktabs}
\usepackage[english]{babel}
\usepackage{amsmath,amssymb,amsbsy,amstext, amsthm, simplewick, amsfonts}
\usepackage{hyperref}
\usepackage{graphicx}
\usepackage[small]{caption}
\usepackage{siunitx}
\usepackage{upgreek}
\usepackage{framed}
\usepackage{wrapfig}
\usepackage{multirow}
\usepackage{bbm}
\usepackage[svgnames,dvipsnames,x11names]{xcolor}
\usepackage{makecell}
\usepackage{selinput}
\usepackage{bm}
\usepackage{float}
\usepackage{geometry}
\usepackage{yfonts}
\usepackage{caption}
\usepackage{subcaption}
\usepackage{sidecap}
\usepackage{longtable}
\usepackage{anyfontsize}
\usepackage{dsfont}
\usepackage{tikz}
\usepackage{relsize}
\usepackage{tcolorbox}
\usetikzlibrary{arrows.meta}
\usetikzlibrary{decorations.pathreplacing,calligraphy}
\usetikzlibrary{decorations.markings, decorations.pathmorphing, decorations.pathreplacing, decorations.shapes}
\usetikzlibrary{shapes.misc}

\usepackage{xcolor}
\usepackage{xparse}

\usepackage{slashed}
\usepackage{simpler-wick}
\usepackage[export]{adjustbox}
\NewDocumentCommand{\colornucleus}{omme{_^}}{%
  \begingroup\colorlet{currcolor}{.}%
  \IfValueTF{#1}
   {\textcolor[#1]{#2}}
   {\textcolor{#2}}
    {%
     #3% the nucleus
     \IfValueT{#4}{_{\textcolor{currcolor}{#4}}}% subscript
     \IfValueT{#5}{^{\textcolor{currcolor}{#5}}}% superscript
    }%
  \endgroup
}

\usepackage{array}
\newcolumntype{L}[1]{>{\raggedright\let\newline\\\arraybackslash\hspace{0pt}}m{#1}}
\newcolumntype{C}[1]{>{\centering\let\newline\\\arraybackslash\hspace{0pt}}m{#1}}
\newcolumntype{R}[1]{>{\raggedleft\let\newline\\\arraybackslash\hspace{0pt}}m{#1}}

\usepackage[framemethod=default]{mdframed}
\newmdenv[skipabove=7pt,
skipbelow=7pt,
rightline=false,
leftline=false,
topline=false,
bottomline=false,
backgroundcolor=gray!10,
linecolor=gray,
innerleftmargin=5pt,
innerrightmargin=5pt,
innertopmargin=5pt,
innerbottommargin=5pt,
leftmargin=0cm,
rightmargin=0cm,
linewidth=4pt]{eBox}
\newmdenv[skipabove=7pt,
skipbelow=7pt,
rightline=false,
leftline=false,
topline=false,
bottomline=false,
backgroundcolor=gray!10,
linecolor=gray,
innerleftmargin=5pt,
innerrightmargin=5pt,
innertopmargin=-5pt,
innerbottommargin=5pt,
leftmargin=0cm,
rightmargin=0cm,
linewidth=4pt]{eBox2}
\usepackage[framemethod=default]{mdframed}
\newmdenv[skipabove=7pt,
skipbelow=7pt,
rightline=true,
leftline=true,
topline=true,
bottomline=true,
backgroundcolor=gray!15,
linecolor=gray,
innerleftmargin=5pt,
innerrightmargin=5pt,
innertopmargin=5pt,
innerbottommargin=5pt,
leftmargin=0cm,
rightmargin=0cm,
linewidth=0.75pt]{eBox3}

\definecolor{Red}{RGB}{214, 39, 40}
\definecolor{Blue}{RGB} {31, 119, 180}
\definecolor{Orange}{RGB}{255, 153, 51}
\definecolor{Purple}{RGB}{178, 102, 255}
\definecolor{Green}{RGB}{44, 160, 44}

\definecolor{vio}{RGB}{19, 130, 164}
\definecolor{vioo}{RGB}{89, 2, 155}
\newcommand{\Comment}[1]{{}}
\definecolor{darkblue}{rgb}{0.15,0.35,0.55}
\definecolor{reddish}{rgb}{0.65, 0.2, 0.2}
\definecolor{darkgreen}{RGB}{50,150,0}
\definecolor{greyish}{rgb}{.90,.90,.90}
\definecolor{greyish2}{rgb}{.96,.96,.96}
\definecolor{greyish3}{rgb}{.37,.37,.37}
\definecolor{darkblue2}{rgb}{0.3,0.4,0.9}
\definecolor{Blue3}{RGB}{31, 119, 180}
\usepackage[linktocpage=true]{hyperref}
\hypersetup{
colorlinks=true,
citecolor=darkblue,
linkcolor=reddish,
urlcolor=darkblue,
pdfauthor={},
pdftitle={},
pdfsubject={}
}

\usepackage{colortbl}
\definecolor{lightgreen}{cmyk}{0.2, 0, 0.2, 0.2}
\definecolor{lightgray2}{cmyk}{0.1,0.1,0,0.1}
\definecolor{Red2}{RGB}{214, 39, 40}
\definecolor{Blue2}{RGB} {31, 119, 180}
\definecolor{Orange2}{RGB}{255, 127, 14}
\definecolor{Green2}{RGB}{44, 160, 44}

\makeatletter
\newlength{\apb@width}
\newcommand{\autoparbox}[2][c]{\settowidth{\apb@width}{#2}\parbox[#1]{\apb@width}{#2}}

\makeatother

\makeatletter
\newlength{\negph@wd}
\DeclareRobustCommand{\negphantom}[1]{%
  \ifmmode
    \mathpalette\negph@math{#1}%
  \else
    \negph@do{#1}%
  \fi
}
\newcommand{\negph@math}[2]{\negph@do{$\m@th#1#2$}}
\newcommand{\negph@do}[1]{%
  \settowidth{\negph@wd}{#1}%
  \hspace*{-\negph@wd}%
}
\makeatother

\def\hs{\hskip 1pt}

\def\beq{\begin{equation}}
\def\eeq{\end{equation}}
\def\be{\begin{equation}}
\def\ee{\end{equation}}

\newcommand{\ud}{{\rm d}}

\definecolor{newred}{RGB}{214, 39, 40}
\definecolor{newblue}{RGB}{31, 119, 180}
\definecolor{neworange}{RGB}{255, 153, 51}

\definecolor{newgreen}{RGB}{44, 160, 44}
\definecolor{newpurple}{RGB}{178, 102, 255}
\definecolor{navy}{RGB}{0, 0, 130}
\definecolor{brown}{RGB}{162, 107, 34}
\definecolor{grey}{RGB}{183, 183, 183}

\def\Lap{{\includegraphics[valign=c]{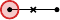}}}
\def\Lam{{\includegraphics[valign=c]{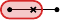}}}
\def\Lbp{{\includegraphics[valign=c]{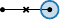}}}
\def\Lbm{{\includegraphics[valign=c]{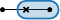}}}

\def\Labg{{\includegraphics[valign=c]{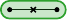}}}
\def\Labr{{\includegraphics[valign=c]{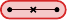}}}
\def\Labb{{\includegraphics[valign=c]{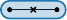}}}

\allowdisplaybreaks[1]
\begin{document}

\newgeometry{top=2cm, bottom=2cm, left=2cm, right=2cm}

\begin{titlepage}
\setcounter{page}{1} \baselineskip=15.5pt 
\thispagestyle{empty}

\begin{center}
{\fontsize{20}{18} \bf Kinematic Flow for Cosmological Loop Integrands}
\end{center}

\vskip 20pt
\begin{center}
\noindent
{\fontsize{14}{18}\selectfont 
Daniel Baumann\hs$^{1,2,3,4}$, Harry Goodhew\hs$^{1,2}$  and Hayden Lee\hs$^{5,6}$}
\end{center}

\begin{center}
  \vskip8pt
\textit{$^1$  Leung Center for Cosmology and Particle Astrophysics,
Taipei 10617, Taiwan}

  \vskip8pt
\textit{$^2$  Center for Theoretical Physics,
National Taiwan University, Taipei 10617, Taiwan}

  \vskip8pt
\textit{$^3$ Institute of Physics, University of Amsterdam, Amsterdam, 1098 XH, The Netherlands}

  \vskip8pt
\textit{$^4$ Max-Planck-Institut f\"ur Physik, Werner-Heisenberg-Institut, 85748 Garching bei M\"unchen, Germany}

\vskip 8pt
\textit{$^5$ Kavli Institute for Cosmological Physics, 
University of Chicago, Chicago, IL 60637, USA}

\vskip 8pt
\textit{$^6$ Department of Physics and Astronomy, University of Pennsylvania, Philadelphia, PA 19104, USA}

\end{center}

%=========================================
\vspace{0.4cm}
\begin{center}{\bf Abstract}
\end{center}
\noindent
Recently, an interesting pattern was found in the differential equations satisfied by the Feynman integrals describing tree-level correlators of conformally coupled scalars in a power-law FRW cosmology~\cite{Arkani-Hamed:2023kig,Arkani-Hamed:2023bsv}. It was proven that simple and universal graphical rules predict the equations for arbitrary graphs as a flow in kinematic space. In this note, we show that the same rules---with one small addition---also determine the differential equations for loop integrands. 
We explain that both the basis of master integrals and the singularities of the differential equations can be represented by tubings of marked graphs. An important novelty in the case of loops is that some basis functions can vanish, and we present a graphical rule to identify these vanishing functions. Taking this into account, we then demonstrate that the kinematic flow correctly predicts the differential equations for all loop integrands.

\end{titlepage}
\restoregeometry

\newpage
\setcounter{tocdepth}{3}
\setcounter{page}{2}

\linespread{1.2}
\tableofcontents
\linespread{1.1}

\newpage
\section{Introduction}

The study of cosmological correlators has both practical and conceptual value. At the practical level, correlation functions are the main observables in cosmology, so computing them to high precision and for a large class of theories plays an important role in the analysis and the interpretation of {\it experimental data}. At a more conceptual level, we are also interested in discovering structural patterns in the theory space of cosmological correlators. 
For this purpose, it is often instructive to study toy models for which cosmological correlators are easier to compute and we can therefore obtain a large amount of {\it mathematical data}. Although these theories are further removed from observations, they allow us to search for hidden patterns that could inform the theory of correlation functions in cosmological spacetimes. A similar philosophy has been pursued for scattering amplitudes where the study of toy models (like supersymmetric Yang--Mills theory) has revealed many fascinating hidden structures (like the amplituhedron). In the context of the cosmological bootstrap~\cite{Snowmass}, toy models have also played an important role in elucidating how physical principles---like locality and unitarity---are reflected in the correlators. It is not unreasonable to expect that some of these insights will eventually also feed back into the structure of real-world cosmological correlators and therefore become of practical relevance.

\vskip 4pt 
Recently, an interesting pattern was discovered in the differential equations satisfied by the Feynman integrals describing tree-level correlators (wavefunction coefficients) of conformally coupled scalars in a power-law FRW cosmology~\cite{Arkani-Hamed:2023kig,Arkani-Hamed:2023bsv}. Up to an overall normalization,
these integrals can be written as
\beq
\psi(X_v, Y_e) = \int \left(\prod_{v=1}^V \ud \omega_v \, \omega_v^{\alpha_v}\right) \psi^{\rm flat}(X_v+\omega_v,Y_e)\,,
\label{equ:twisted}
\eeq
where $X_v$ and $Y_e$ are the external and internal energies of a Feynman graph, respectively, and $\alpha_v$ are theory-dependent constant parameters.
The function $\psi^{\rm flat}(X_v+\omega_v,Y_e)$ in the integrand is the flat-space correlator (wavefunction coefficient), which is a rational function~\cite{Arkani-Hamed:2017fdk}. Integrals of the form~(\ref{equ:twisted}) are called ``twisted integrals," with $\omega_v^{\alpha_v}$ being twist factors. It can be shown that such integrals are part of a finite-dimensional vector space of master integrals, $\vec{I} \equiv [\psi, I_2, \cdots, I_N]^T$, and that
taking derivatives with respect to the kinematic variables $Z_I \equiv (X_v,Y_e)$ leads to coupled differential equations for these integrals:
\beq
{\rm d} \vec{I} =  A \vec{I}\,,
\eeq
where ${\rm d} \equiv \sum {\rm d} Z_I \partial_{Z_I}$
and $A$ is an $N \times N$ ``connection matrix". Explicitly,
this connection matrix can be written as a sum of dlog forms 
\beq
A = \sum_i A_i\, \ud\hskip -1pt \log \Phi_i(Z)\,,
\eeq 
where $A_i$ are constant matrices and the functions $\Phi_i(Z)$ are called ``letters”.

\vskip 4pt
At first sight, the matrices $A$ look very complicated and the pattern of letters appears random. Something remarkable, however, happens when this information is represented graphically. Both the letters and the basis functions can be described by ``tubings" of marked graphs, and very simple rules---called the ``kinematic flow"---capture how these elements appear in the differential equations~\cite{Arkani-Hamed:2023kig,Arkani-Hamed:2023bsv}. This revealed a hidden order in the apparent randomness. So far, the rules of the kinematic flow have only been established for tree graphs.  In this note, we show that the same rules---with one small addition---also determine the differential equations for loop integrands. 

\paragraph{Online} The outline of this paper is as follows: In Section~\ref{sec:review}, we review the toy model of conformally coupled scalars in a power-law FRW cosmology.  We show that the wavefunction coefficients in this theory can be written as twisted integrals over the energy variables, and describe the differential equation method for studying such integrals. In Section~\ref{sec:KinematicFlow}, we derive the differential equations for one- and two-loop integrands. We show that these equation obey the rules of the kinematic flow, if we take into account that some of the basis functions can vanish identically. We present a simple graphical rule for identifying such vanishing functions a priori. Finally, in Section~\ref{sec:Conclusions}, we state our conclusions and present a list of open problems.

\vskip 10pt
\noindent
{\it Note added:} While this paper was in preparation, ref.~\cite{He:2024olr} appeared which also studied the differential equations for cosmological loop integrands using a different method (see also~\cite{Fan:2024iek}). %Where comparison is possible, we agree with their results. In addition, we show that the results can be obtained  graphically. 
We agree with their results and show that they can be obtained graphically.  
As we prepared our arXiv submission, ref.~\cite{Hang} appeared with the kinematic flow for the one-loop bubble. Here, we present flow rules that apply to arbitrary loops.

\section{Correlators as Twisted Integrals}
\label{sec:review}

Like in \cite{Arkani-Hamed:2023kig}, we will consider conformally coupled scalars in a power-law FRW universe with polynomial interactions.
We will study loop integrals in dimensional regularization and therefore keep the spacetime dimension $D$ arbitrary.
The action for this toy model then is
\begin{align}
  S= \int {\rm d}^D x\, \sqrt{-g}\left[-\frac{1}{2}\partial_{\mu}\phi\partial^{\mu}\phi-\frac{D-2}{8(D-1)}R\phi^2-\sum_{p=3}^\infty \frac{\lambda_p}{p!} \phi^p\right] ,
\end{align}
where $R$ is the Ricci scalar.
We will assume that the scale factor takes the form of a power law $a(\eta)= (\eta/\eta_0)^{-(1+\varepsilon)}$, where $\eta $ is conformal time that runs from $-\infty$ to $0$, and $\varepsilon$ is a constant parameter. The values $\varepsilon=0,-1,-2,-3$ correspond to de Sitter, flat, radiation-dominated, and matter-dominated universes, respectively. For simplicity, we will work in units where  
$\eta_0 \equiv -1$. 
The reason for choosing this toy model is that its mode function is simple, $\phi_k(\eta) = (-\eta)^\beta \phi_k^{\rm flat}(\eta)$, where $\beta \equiv \frac{1}{2}(D-2)(1+\varepsilon)$ and $\phi_k^{\rm flat}(\eta) = e^{ik\eta}/\sqrt{2k}$. The model is still nontrivial, since generic polynomial interactions are non-conformal and are sensitive to the background spacetime.

\vskip 4pt
In this section, we will show that the wavefunction coefficients in this theory can be written as twisted integrals of the corresponding flat-space results~\cite{Arkani-Hamed:2017fdk}.
This furthermore implies that the wavefunction coefficients are part of a finite-dimensional vector space of master integrals, which can be studied using the method of differential equations~\cite{Bern:1993kr,Kotikov:1990kg,Remiddi:1997ny,Gehrmann:1999as,Henn:2013pwa,Henn:2014qga,Abreu:2022mfk}.

\subsection{Wavefunction Coefficients}\label{sec:wf}

We are interested in correlation functions at a fixed time $\eta=\eta_*$, which can be written as the following functional integral
\beq
\langle \varphi({\bf x}_1) \ldots \varphi({\bf x}_N) \rangle = \int {\cal D} \varphi\ \varphi({\bf x}_1) \ldots \varphi({\bf x}_N)\, |\Psi[\varphi]|^2\,,
\eeq
where $\varphi({\bf x}) $ is the boundary value of the field $\phi(\eta,{\bf x})$. In this work, we will be concerned with the analytic structure of the wavefunction $\Psi[\varphi]$. For small fluctuations, we expand the wavefunction (in Fourier space) as
\begin{align}
    \Psi[\varphi]&=\exp\left[-\sum_{n=2}^\infty \int \left(\prod_{a=1}^n\frac{\ud^dk_a}{(2\pi)^d}\varphi_{\textbf{k}_a}\right)\psi_n(\textbf{k}_1,\dots,\textbf{k}_n) (2\pi)^d \delta^d
    \left(\textbf{k}_1+\dots+\textbf{k}_n\right)\right],
\end{align}
where $d=D-1$ is the spatial dimension.
The kernel functions $\psi_n$ are the  ``wavefunction coefficients".
They can be computed using the following Feynman rules:\\[4pt]
$\bullet$ Draw all graphs with $n$ lines ending on the spatial surface at $\eta=\eta_*\approx 0$:\footnote{More precisely, we only keep the leading terms in the limit $\eta_*\to 0$, so that $\phi_k(\eta_*) = (-\eta_*)^\beta /\sqrt{2k}$.} 
 \begin{equation}
 \raisebox{-26pt}{
\begin{tikzpicture}[line width=1. pt, scale=2]
\draw[fill=black] (0,0) -- (1,0);
\draw[line width=1.pt,lightgray] (0,0) -- (-0.25,0.55);
\draw[line width=1.pt,lightgray] (0,0) -- (0,0.55);
\draw[line width=1.pt,lightgray] (0,0) -- (0.25,0.55);
\draw[line width=1.pt,lightgray] (1,0) -- (0.75,0.55);
\draw[line width=1.pt,lightgray] (1,0) -- (1,0.55);
\draw[line width=1.pt,lightgray] (1,0) -- (1.25,0.55);
\draw[lightgray, line width=2.pt] (-0.5,0.55) -- (1.5,0.55);
\draw[fill=Red,Red] (0,0) circle (.03cm);
\draw[fill=Blue,Blue] (1,0) circle (.03cm);
\node[scale=1] at (0,-.15) {\small $X_1$};
\node[scale=1] at (1,-.15) {\small $X_2$};
\node[scale=1] at (0.5,-.12) {\small $Y$};
\node[scale=1] at (-0.25,0.67) {\small $k_1$};
\node[scale=1] at (0,0.67) {\small $\cdots$};
%\node[scale=1] at (0.25,0.67) {\small $k_3$};
%\node[scale=1] at (0.75,0.67) {\small $k_4$};
\node[scale=1] at (1,0.67) {\small $\cdots$};
\node[scale=1] at (1.25,0.67) {\small $k_n$};

\begin{scope}[xshift=3cm]
%\draw[fill=black] (0,0) -- (1,0);
\draw (0,0) to [bend right] (1,0);
\draw (0,0) to [bend left] (1,0);
\draw[line width=1.pt,lightgray] (0,0) -- (-0.25,0.55);
\draw[line width=1.pt,lightgray] (0,0) -- (0,0.55);
\draw[line width=1.pt,lightgray] (0,0) -- (0.25,0.55);
\draw[line width=1.pt,lightgray] (1,0) -- (0.75,0.55);
\draw[line width=1.pt,lightgray] (1,0) -- (1,0.55);
\draw[line width=1.pt,lightgray] (1,0) -- (1.25,0.55);
\draw[lightgray, line width=2.pt] (-0.5,0.55) -- (1.5,0.55);
\draw[fill=Red,Red] (0,0) circle (.03cm);
\draw[fill=Blue,Blue] (1,0) circle (.03cm);
\node[scale=1] at (0,-.15) {\small $X_1$};
\node[scale=1] at (1,-.15) {\small $X_2$};
\node[scale=1] at (0.5,-.27) {\small $Y_2$};
\node[scale=1] at (0.5,.27) {\small $Y_1$};
\node[scale=1] at (-0.25,0.67) {\small $k_1$};
\node[scale=1] at (0,0.67) {\small $\cdots$};
%\node[scale=1] at (0.25,0.67) {\small $k_3$};
%\node[scale=1] at (0.75,0.67) {\small $k_4$};
\node[scale=1] at (1,0.67) {\small $\cdots$};
\node[scale=1] at (1.25,0.67) {\small $k_n$};
\end{scope}
\end{tikzpicture}
} 
\nonumber
\end{equation}
$\bullet$ Assign a bulk-to-boundary propagator to each external line
\beq
K_k(\eta) = \frac{\phi_k(\eta)}{\phi_k(\eta_*)} = \left(\frac{\eta}{\eta_*}\right)^\beta e^{ik(\eta-\eta_*)}\,,
\eeq
where $\phi_k(\eta)$ is the solution of the free-field equation.\\[4pt]
$\bullet$ Assign a bulk-to-bulk propagator to each internal line
\beq
G_k(\eta,\eta') = \phi_k^*(\eta)\phi_k(\eta')\, \theta(\eta-\eta')+ \phi_k^*(\eta')\phi_k(\eta) \,\theta(\eta'-\eta) - \frac{\phi_k^*(\eta_*)}{\phi_k(\eta_*)}\phi_k(\eta)\phi_k(\eta')\,, \label{equ:G}
\eeq
where $\theta(\eta-\eta')$ is the Heaviside function enforcing time-ordering.\\[4pt]
$\bullet$ Assign a factor of $i\lambda_p$ to each bulk vertex.\\[4pt]
$\bullet$ Integrate over all vertex times, using $\ud\eta\, a^{d+1}(\eta)$, and over any undeterminded loop momenta.

\vskip 4pt
The wavefunction coefficients for a graph with $V$ vertices and $L$ loops then takes the form 
\begin{align}\label{eq:WFn}
    \psi_n(\{\textbf{k}_a\})&=\int \left(\prod_{v=1}^V \ud\eta_v\, a^{d+1}(\eta_v)(i\lambda_v)\prod_{a=1}^{n_v} K_{k_a}(\eta_v) \right)\int \prod_{l=1}^L \ud^d \textbf{Y}_l \prod_{i=e}^E G_{Y_e}(\eta_e,\eta_e')\,,
\end{align}
where $\lambda_v$ is the coupling constant for the interaction of the vertex at $\eta_v$, $n\equiv\sum_v n_v$ and $\textbf{Y}_l$ are the loop momenta. This expression holds only for theories involving no derivatives. 
It is straightforward to generalize to the case where they are of interest, but we will not be considering such theories here. 
We will start by considering tree-level wavefunction coefficients in arbitrary power-law FRW spacetimes and show how they can be written as integrals over the corresponding flat-space results (which are easy to derive). 
The generalization of these relationships to loop integrands will be given in the next section.

\vskip 4pt
An important aspect of the mode functions for conformally coupled scalars is that they are proportional to those in flat space, %\HL{cf.~\eqref{eq:mode}.} 
$\phi_k(\eta)=(-\eta)^{\beta} \phi_k^{\text{flat}}(\eta)$. 
Inserting this solution into our expression~\eqref{eq:WFn} for the wavefunction coefficients, we have 
\begin{equation}
  \psi_n^{\rm tree}(\{\textbf{k}_a\}) =  (-\eta_*)^{-\beta n } \int \left(\prod_{v=1}^V \frac{\ud\eta_v}{(-\eta_v)^{1+\alpha_v}} (i\lambda_v)\prod_{a=1}^{n_v} K^{\text{flat}}_{k_a}(\eta_v)\right)\prod_{e=1}^E G^{\text{flat}}_{Y_e}(\eta_e,\eta_e')\,,
  \label{equ:TREE}
\end{equation}
where  $\alpha_v \equiv d+\varepsilon(d+1)+\frac{1}{2}p_v(1+\varepsilon)(1-d)$ and $p_v$ is the number of lines entering the vertex~$v$. For $d=3$, we have $\alpha_v = (3-p_v)+(4-p_v) \varepsilon$, which reduces to $\alpha_v = \varepsilon$ for cubic interactions. Using the identity
\beq
\frac{1}{(-\eta)^{1+\alpha}} = \frac{ie^{\frac{i\pi \alpha}{2}}}{\Gamma(1+\alpha)}\int_0^\infty \ud\omega \,\omega^\alpha e^{i\omega\eta}\,,
\eeq
we can write (\ref{equ:TREE}) as
\begin{align}
    \psi_n^{\rm tree}(\{\textbf{k}_a\})
    &\propto \int \left(\prod_{v=1}^V \ud\omega_v\, \omega_v^{\alpha_v}\right) \psi_{n,\text{tree}}^{\text{flat}}(X_v+\omega_v,Y_e)\,,
\end{align}
where $X_v$ is the sum of the external energies at vertex $v$, and we have suppressed the overall prefactor which depends on $\alpha_v$ and $\eta_*$. We see that the tree-level wavefunction coefficients in FRW spacetimes are generated by a frequency integral for each vertex over the flat-space wavefunction coefficient.  
For later convenience, we will work with the rescaled wavefunction coefficient 
\begin{equation}
    \psi_{(V)}^{\rm tree}(X_v,Y_e)
    =  \int \left(\prod_{v=1}^V \ud\omega_v\, \omega_v^{\alpha_v}\right)\left(\prod_{e=1}^E 2Y_e\right)\psi_{(V),\text{tree}}^{\text{flat}}(X_v+\omega_v,Y_e)\,, \label{eq:Tree}
\end{equation}
where we have introduced a factor of $2Y_e$ for every internal line. We will later absorb these factors into the definition of the flat-space wavefunction coefficients. Since the wavefunction coefficient only depends on the total energies that enter the individual vertices, independent of the number of external legs, we have labelled it with the subscript $V$.
%As we will explain shortly, the additional factors of internal energies allow us to interpret the integrand as the volume of a geometry. 

\subsection{Loops in Dimensional Regularization}
We would now like to extend the discussion to wavefunction coefficients involving loops. The relevant loop integrals can be written as 
\begin{align}\label{eq:phihat}
    \psi_n^{\rm loop}(\{\textbf{k}_a\}) =\int\prod_{l=1}^{N_L}\ud^d  \textbf{Y}_l\,\hat{\psi}_n^{\rm loop}(\{\textbf{k}_a\}) \,. 
\end{align}
Performing these loop integrals is challenging, mostly due to nontrivial measures of integration; see~\cite{Benincasa:2024lxe, Benincasa:2024ptf}
for recent progress.
In this paper, we will therefore focus on the loop integrand $\hat{\psi}_n^{\rm loop}(\{\textbf{k}_a\})$.
Using the Feynman rules, it is given by 
\begin{equation}
  \hat\psi_n^{\rm loop}(\{\textbf{k}_a\}) =  (-\eta_*)^{-\beta n } \int \left(\prod_{v=1}^V \frac{\ud\eta_v}{(-\eta_v)^{1+\alpha_v}} (i\lambda_v)\prod_{a=1}^{n_v} K^{\text{flat}}_{k_a}(\eta_v)\right)\prod_{e=1}^E G^{\text{flat}}_{Y_e}(\eta_e,\eta_e')\,,
\end{equation}
which has the same form as \eqref{equ:TREE}. 
Just like in the tree-level case, we can write this as a twisted integral over the flat-space wavefunction coefficient:
\begin{align}
    \hat \psi_{(V)}^{\rm loop}(X_v,Y_e)
    &= \int \left(\prod_{v=1}^V \ud\omega_v\, \omega_v^{\alpha_v}\right)\left(\prod_{e=1}^E 2Y_e\right)\hat\psi_{(V),\text{loop}}^{\text{flat}}(X_v+\omega_v,Y_e)\,,
    \label{eq:Loop}
\end{align}
where we have have introduced factors of $2Y_e$ for every internal lines as in \eqref{eq:Tree}. 
We see that the expression (\ref{eq:Loop}) is of the same form as \eqref{eq:Tree} and the only difference is in the topology of the graph. 
In the following, we will drop the hat on $\hat \psi^{\rm loop}_{(V)}$, with the understanding that we will always be working with the loop integrand throughout the rest of the paper.

\subsection{Integrands from Graph Tubings}
\label{ssec:tubings}

Although the flat-space wavefunction can be computed using the Feynman rules outlined in Section~\ref{sec:wf}, higher-order calculations become increasingly complex due to the large number of terms needed to be kept with various time orderings. However, an important insight from \cite{Arkani-Hamed:2017fdk} is that the flat-space wavefunction admits an intrinsically combinatorial definition, which circumvents the need to introduce time integrals.
To reveal this, it is useful to represent these wavefunctions graphically in terms of ``graph tubings". 

\vskip 4pt
First of all, the result for the flat-space wavefunction depends only on the total energy
 flowing into each vertex. %This is true for both trees and loops, as shown in \eqref{eq:Tree} and \eqref{eq:Loop}.
We can therefore simply represent a Feynman diagram by a graph with all external lines truncated, labeling each vertex with its associated total external energy.
Given such a graph, there is a simple combinatorial procedure that outputs the corresponding flat-space wavefunction. 

\vskip 4pt
This procedure requires decomposing the graph into subgraphs, which we represent graphically by encircling each subgraph with a ``tube".
To each tube, we associate the sum of external and internal energies enclosed by it.  Finally, a ``complete tubing" is a maximal set of non-overlapping tubes of a graph. To compute the flat-space wavefunction coefficients, we first generate the set of all compatible
complete tubings of a given graph. We then assign to each complete tubing the inverse of the product of all of the energies associated to the
constituent tubings, and sum over all possible compatible tubings of the graph.
The result is a remarkably simple formula:
\beq\label{eq:FlatWFC}
\psi^{\rm flat} = -\sum_{\cal T} \prod_a \frac{1}{E_a}\,,
\eeq
where ${\cal T}$ is the set of all compatible (non-overlapping) complete tubings of the graph, $E_a$ denotes the total energy associated to each tube, and
we have set the coupling constants to unity ($\lambda_p \equiv 1$).
The above formula is slightly abstract, so let us illustrate it with two simple examples.

\paragraph{Two-site chain} The simplest tree-level example is the two-site chain. In that case, there is a unique complete tubing and the associated wavefunction coefficient is 
\be\label{equ:2pt-tubing}
 \psi_{(2),{\rm tree}}^{\rm flat} \ = \  \includegraphics[valign=c]{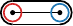}
\,\,=\,\,  -\frac{2Y}{{\color{black}(X_1+X_2)}{\color{newred}(X_1+Y)}{\color{newblue}(X_2+Y)}} \,.
\ee
%This function has been colour coded according to which tube gives which term in the denominator. 
%Furthermore, as mentioned in the previous section this function 
%\hg{we also mention this above, this is the difference between $\hat{\psi}$ and $\psi$} 
As discussed in~\cite{Arkani-Hamed:2017fdk}, we have included an overall factor of $2Y$, so that  this wavefunction coefficient can be associated to the ``canonical form" of the region enclosed by the three lines
\beq
\begin{aligned}
    {\color{newred} B_1} &=X_1+Y+\omega_1\,,\\ {\color{newblue} B_2} &=X_2+Y+\omega_2\,, \\ B_3 &=X_1+X_2+\omega_1+\omega_2\,,
    \end{aligned}
    \label{Btwositechain}
\eeq
which is
\beq
\Omega_{123} \equiv \ud \log \left(\frac{B_1}{B_3}\right) \wedge \ud \log\left(\frac{B_2}{B_3}\right) =- \frac{2Y}{B_1 B_2 B_3} \hs 
 \ud \omega_1 \wedge \ud \omega_2\,.
 \label{equ:Omega-chain}
\eeq
Relating the wavefunction to a canonical form is not only natural from a geometric point of view, but it also proves very useful for providing a canonical representation for the differential equations it satisfies. 

\paragraph{One-loop bubble} Our next example is the one-loop bubble diagram. In that case, there are two inequivalent complete tubings and the associated wavefunction coefficient is therefore the sum of two contributions 
    \begin{align}
&\psi^{\text{flat}}_{(2),\text{bubble}}=\ \includegraphics[valign=c]{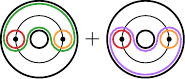} \nonumber\\
&\quad =-\frac{4Y_1Y_2}{\color{newred}{(X_1+Y_1+Y_2)}\color{neworange}{(X_2+Y_1+Y_2)}\color{black} (X_1+X_2)}\left(\frac{1}{\color{newgreen}X_1+X_2+2Y_2}\color{black}+\frac{1}{\color{newpurple}X_1+X_2+2 Y_1}\color{black}\right).\label{flatbubble}
    \end{align}
 Again, this function has a geometrical interpretation as the canonical form of a region bounded by the 
  lines 
\begin{equation}
    \begin{aligned}
        \color{newred}B_1&=X_1+Y_1+Y_2+\omega_1\,,&\color{neworange}B_2&=X_1+Y_1+Y_2+\omega_2\,,\\B_3&=X_1+X_2+\omega_1+\omega_2\,,&
        \color{newgreen}B_4&=X_1+X_2+2Y_2+\omega_1+\omega_2\,,\\\color{newpurple}B_5&=X_1+X_2+2Y_1+\omega_1+\omega_2\,,
    \end{aligned}
\end{equation}
which is 
\beq
\begin{aligned}
    {\Omega}_{12345}&={\Omega}_{123}-{\Omega}_{124}-{\Omega}_{125}=-\frac{4Y_1Y_2}{\color{newred}{B_1}\color{neworange}{B_2}\color{black}B_3}\left(\frac{1}{\color{newgreen}B_4}\color{black}+\frac{1}{\color{newpurple}B_5}\color{black}\right) \ud\omega_1\wedge\ud\omega_2 \,.
    %\\&=-[B_2B_3]-[B_3B_1]+[B_2B_5]+[B_5B_1]+[B_1B_2]+[B_2B_4]+[B_4B_1]\,.
\end{aligned}
\label{equ:Omega-bubble}
\eeq
As we have indicated, the canonical form can be obtained by triangulating it into three simplices. We will see shortly that defining the wavefunction integrand in terms of the canonical forms of these simplices is very useful for deriving the corresponding differential equations. 

\subsection{Differential Equations}

We have seen that both the tree-level wavefunction coefficients and the loop integrands can be written as twisted integrals of the form
\beq
\psi_{(V)}(X_v) = \int \left(\prod_{v=1}^V \ud\omega_v\, \omega_v^{\alpha_v}\right) \psi_{(V)}^{\rm flat}(X_v+\omega_v,Y_e)\,.
\eeq
As explained in~\cite{Arkani-Hamed:2023kig}, such integrals are part of a finite-dimensional vector space of master integrals:
\beq
\vec I \equiv  \left[ \begin{array}{c} \psi \\ I_2 \\ I_3 \\ \vdots \\ I_N \end{array} \right] .
\label{equ:family}
\eeq
Taking derivatives with respect to the kinematic variables then leads to coupled differential equations for these integrals. Defining the total differential ${\rm d} \equiv \sum {\rm d} Z_I \partial_{Z_I}$, where $Z_I$ contains both the external energies $X_v$ and the internal energies $Y_e$, we get
\beq
{\rm d} \vec{I} =  A \vec{I}\,,
\eeq
where $A$ is an $N \times N$ ``connection matrix" whose entries are one-forms. Explicitly,
this connection matrix can be written as a sum of dlog-forms 
\beq
A = \sum_i A_i\, \ud \log \Phi_i(Z)\,,
\eeq 
where $A_i$ are constant matrices and the functions $\Phi_i(Z)$ are called ``letters''.
The set of all letters is the ``alphabet" and it determines the possible singularities of functions in the family~\eqref{equ:family}. 

\vskip 4pt
As we discussed above, the (loop integrand of) the flat-space wavefunction is a rational function of the energies. 
When multiplied by the twist factor, $\prod_v\omega_v^{\alpha_v}$, we can study the vector space of these integrals using the ``twisted cohomology'' of the ``hyperplane arrangements''. 
The number of bounded chambers carved out by the hyperplanes then determines the dimension of the vector space for a given family of generic twisted integrals. 
While~\cite{Arkani-Hamed:2023kig} found that the hyperplane arrangements for tree-level wavefunction coefficients are highly non-generic and obey smaller differential systems, we will show that the counting of bounded chambers precisely matches the dimension of the vector space of the loop integrands. 

\vskip 4pt
In order to write the connection matrix in terms of dlog forms, it is necessary to make an appropriate choice of master integrals. This can be achieved by defining them using canonical forms of simplices. In particular, it is useful to introduce projective simplices as done in \cite{Arkani-Hamed:2023kig}, which are simplices formed with the plane at infinity. 
In $n$ dimensions, the canonical forms of these projective simplices, or projective simplex forms, are given by 
\begin{equation}
    [L_1\cdots L_n] \equiv \ud\log L_1\wedge\cdots \wedge \ud\log L_n\,,
\end{equation}
where $L_I$ represent the hyperplanes involved, and the exterior derivatives are with respect to the internal variables $\omega_a$. 
The boundary of a simplex form is defined as
\begin{align}
    \partial[L_1\cdots L_n] = \sum_{J=1}^n(-1)^{J+1}[L_1\cdots \widehat L_J\cdots L_n]\,,
\end{align}
where $\,\widehat{\phantom{.}}\,$ indicates an omission. 
Linear combinations of these forms can be used to construct bounded regions involving the lines defined by the energy singularities of a given wavefunction coefficient as well as the coordinate axes that define the twist in the integral. 
These canonical forms will be our master integrals.

\paragraph{Two-site chain} 

Before considering loop diagrams, we begin with a brief review of the two-site chain, which represents the simplest nontrivial diagram.
The differential equations for the two-site chain were derived in \cite{De:2023xue, Arkani-Hamed:2023kig}.

\vskip 4pt
The two-site chain has four master integrals, which were chosen in \cite{Arkani-Hamed:2023kig} as
\begin{align}
\left[\begin{array}{c}
     \psi  \\ F \\ \tilde F \\ Z
\end{array}\right] = \int \omega_1^{\alpha_1}\omega_2^{\alpha_2}\left[\begin{array}{c}
     \Omega_\psi  \\ \Omega_F \\ \Omega_{\tilde F} \\ \Omega_Z
\end{array}\right],
\end{align}
with 
\beq
\begin{aligned}
\Omega_\psi &= \partial[B_1B_2B_3] =  \frac{-2Y}{B_1 B_2 B_3} \, \ud \omega_1 \wedge \ud \omega_2 \,,\quad \Omega_{\tilde F} = \partial[B_1T_2B_3] =  \frac{X_2-Y}{B_1 T_2 B_3} \, \ud \omega_1 \wedge \ud \omega_2\,,\\[4pt]
\Omega_F &= \partial[T_1B_2B_3] =  \frac{X_1-Y}{T_1 B_2 B_3} \, \ud \omega_1 \wedge \ud \omega_2 \,,\quad\ \Omega_Z =\partial[T_1T_2B_3]=  \frac{X_1+X_2}{T_1 T_2 B_3}\,\ud \omega_1 \wedge \ud \omega_2\,,
\end{aligned}
\eeq
where $T_i \equiv \omega_i$ are the ``twisted lines" and the ``boundary lines" $B_i$ were defined in \eqref{Btwositechain}. 
The differential equations take the form\footnote{This set of differential equations forms a restricted Gelfand-Kapranov-Zelevinsky (GKZ) system~\cite{Gelfand:1990bua}, which can be decomposed into smaller subsystems~\cite{Grimm:2024tbg,Fevola:2024nzj}.}
\begin{align}
&\ud \psi = \alpha_1 \, \big[(\psi - F)\,\ud \log X_1^+ + F\, \ud \log X_1^-\big] +  \alpha_2\,\big[(\psi - \tilde F)\, \ud \log X_2^+ +\tilde F\,\ud \log X_2^- \big]\,,\nonumber \\[4pt]
\begin{split}
\ud F &= \alpha_1\, \big[ F\,\ud \log X_1^-  + (F-  Z)\, \ud \log X_2^+ + Z\, \ud \log X_{12} \big] \,,  \\[4pt]
\ud \tilde F &=  \alpha_2\, \big[ \tilde F\,\ud \log X_2^-  + (\tilde F-  Z)\, \ud \log X_1^+ + Z\, \ud \log X_{12} \big]\,,  \\[4pt]
\end{split} \label{eq:twositechain}\\
&\ud Z = 2 (\alpha_1+\alpha_2)\, Z\, \ud \log X_{12}\,,\nonumber
\end{align}
where $X_i^\pm \equiv X_i \pm Y$ and $X_{12}\equiv X_1+X_2$.

\paragraph{One-loop bubble}

Let us now present the differential equations for the one-loop bubble. 
For this purpose, it is convenient to consider a different decomposition of the bubble than the one shown in \eqref{flatbubble}.
Using the tree theorem~\cite{AguiSalcedo:2023nds}, we can decompose the bubble integrand into a sum of three two-site chains:
\begin{align}
    \psi_{(2)}^{\text{bubble}}= \psi^{\rm I}_{(2)}+  \psi^{\rm II}_{(2)}+ \psi^{\rm III}_{(2)}\,,\label{eq:bubbletreedecomp}
\end{align}
where 
\beq
\begin{aligned}
    \psi_{(2)}^{\rm I}&= \psi_{(2)}^{\text{tree}}(X_1+Y_1,X_2+Y_1,Y_2)\,,\\
    \psi_{(2)}^{\rm II}&=-\psi_{(2)}^{\text{tree}}(X_1,X_2,Y_1+Y_2)\,,\\
    \psi_{(2)}^{\rm III}&= \psi_{(2)}^{\text{tree}}(X_1+Y_2,X_2+Y_2,Y_1)\,.
\end{aligned}
\eeq
Each decomposed tree diagram then satisfies the $4\times 4$ system of differential equations given in~\eqref{eq:twositechain}. 
This means that each tree-level diagram gives 3 additional functions, resulting in a total of $1+3\times 3=10$ functions for the one-loop bubble.

\vskip 4pt
It is then straightforward to write down the differential equations for the bubble by combining three sets of tree-level equations. 
We choose the basis vector as 
\begin{equation}
    \vec I=[\psi,F^{(\text{I})},F^{(\text{II})},F^{(\text{III})},\tilde F^{(\text{I})},\tilde F^{(\text{II})},\tilde F^{(\text{III})},Z^{(\text{I})},Z^{(\text{II})},Z^{(\text{III})}]^T\,,
    \label{equ:Bubble-Basis}
\end{equation}
where $F^{(i)},\tilde F^{(i)},Z^{(i)}$ denote the three basis functions associated to the tree-level terms in \eqref{eq:bubbletreedecomp}. 
For simplicity, let us show the result for the equal-twist case $\alpha_1=\alpha_2=\varepsilon$, for which the system of differential equations can be written as
\begin{align}\label{eq:dI}
	\ud \vec I = \varepsilon A\vec I\,,
\end{align}
where $A$ is a matrix-valued dlog form
\begin{align}
	A = \ud\log\left[\begin{array}{cccccccccc}
 \ell_1\ell_5 & \ell_4/\ell_1 & \ell_3/\ell_1 & \ell_2/\ell_1& \ell_8/\ell_5& \ell_7/\ell_5&\ell_6/\ell_5 &0 & 0&	0\\ 
 0&\ell_4\ell_5 &0 &0 &0 &0 &0 &\ell_{11}/\ell_5 &0 &0\\
  0& 0& \ell_3\ell_5& 0& 0& 0& 0& 0& \ell_9/\ell_5&0\\
  0 &0 &0 &\ell_2\ell_5 &0 &0 &0 &0 &0 &\ell_{10}\ell_5\\
   0 &0 &0 &0 &\ell_1\ell_8 &0 &0 &\ell_{11}/\ell_1 &0 &0\\
     0& 0& 0& 0& 0& \ell_1\ell_7& 0& 0& \ell_9/\ell_1&0\\
     0 &0 &0 &0 &0 &0 &\ell_1\ell_6 &0 &0 &\ell_{10}/\ell_1\\
      0& 0& 0& 0& 0& 0& 0& 2\ell_{11}&0 &0\\
     0 & 0& 0& 0& 0& 0& 0& 0& 2\ell_9&0\\            
      0 & 0& 0& 0& 0& 0& 0& 0& 0&2\ell_{10}\\
 \end{array}\right],
 \label{equ:bubble}
\end{align}
where the letters $\ell_i$ are 
\beq
\begin{aligned}
	\ell_1 &= X_1^{++}\,,\quad \ell_2 = X_1^{-+}\,,\quad \ell_3 = X_1^{+-}\,,\quad \ell_4 = X_1^{--}\,,\\
	\ell_5 &= X_2^{++}\,,\quad \ell_6 = X_2^{-+}\,,\quad \ell_7 = X_2^{+-}\,,\quad \ell_8 = X_2^{--}\,,\\
	\ell_{9} &= X_{12}^{+0}\,,\quad \ell_{10} = X_{12}^{0+}\,,\quad \ell_{11} = X_{12}^{00}\,,\label{letters1}
\end{aligned}
\eeq
with
\beq
\begin{aligned}
	X_i^{ab} \equiv X_i +a Y_1 +b Y_2\,,\quad X_{12}^{ab} \equiv X_1+X_2 + 2a Y_1 +2b Y_2\,.
\end{aligned}
\eeq
 This matrix was also found in \cite{De:2023xue} using a different method.

\vskip 4pt 
It is straightforward to solve the system of differential equations defined by (\ref{equ:bubble}). In fact, since the bubble can be written as a sum of tree-level terms, we could immediately write down the answer using the tree-level solution derived in \cite{Arkani-Hamed:2023kig}. 
In the de Sitter limit, $\varepsilon \to 0$, there is an alternative way of finding the solution using the {\it symbol}. 
The loop integrands are expressed in terms of transcendental functions; these functions take the form of iterated integrals, whose structure is elegantly encoded in the symbol. The symbol fully determines the transcendental function up to a rational part, which for the cosmological loop integrands can be easily fixed by imposing the regularity of the solution~\cite{Hillman:2019wgh}.

\paragraph{General graphs} In general, the matrices $A$, and hence the corresponding differential equations, can be very complex. It was therefore rather remarkable when it was found in~\cite{Arkani-Hamed:2023kig} that the equations for all tree-level graphs can be predicted using simple and universal graphical rules called the ``kinematic flow".
In the next section, we will show that the same rules---with one small addition---also lead to the differential equations for the loop integrands.

\section{Kinematic Flow for Loop Integrands}
\label{sec:KinematicFlow}

We are now ready to discuss the graphical representation of the differential equations for loop integrands. As in~\cite{Arkani-Hamed:2023kig}, the letters and basis functions of the differential equations will be characterized by tubings of marked graphs.
We will demonstrate explicitly that the pattern by which these elements appear in the differential equations for arbitrary loop graphs can be predicted by the rules of the ``kinematic flow".

\subsection{Letters and Marked Graphs}

As can be seen in (\ref{eq:twositechain}) and (\ref{equ:bubble}), the differential equations for the FRW wavefunction coefficients contain the same energy singularities as the flat-space wavefunction, which we represented as graph tubings in Section~\ref{ssec:tubings}. In addition, it has singularities with flipped signs of the internal energies. To represent the complete set of singularities (``letters") graphically, we follow \cite{Arkani-Hamed:2023kig} and introduce the concept of ``marked graphs". Each internal line is marked by a cross and when a graph tubing encloses the cross it signifies a sign flip of the corresponding internal energy.
For example, the equations for the two-site chain in  (\ref{eq:twositechain}) can be written graphically as 
\begin{align}
\ud \psi
&\ =  \  \alpha_1\, \Big[(\psi-F)
\  \Lap
 \ + \   F
\  \Lam \Big]
%%%%%%%%%%%%%%%%%%%%%%%%%%%%%%%%%%
 \ + \   \alpha_2\, \Big[(\psi- \tilde F)
\  \Lbp 
\ \ + \ \  \tilde F
\   \Lbm \Big]\,, \label{equ:TwoSite-dPsi} \\[-4pt]
\cline{1-2}
%%%%%%%%%%%%%%%%%%%%%%%%%%%%%
\ud F 
&\ =  \   \alpha_1\, \Big[ F 
\  \Lam 
 \ +  \  (F-Z)\ \Lbp 
\ + \ 
Z\  \Labb \Big]\,, \label{equ:TwoSite-dF-X}
  \\[5pt]
 %%%%%%%%%%%%%%%%%%%%%%%%%%%%%
\ud \tilde F 
&\ =\  \alpha_2\, \Big[ \tilde F 
\ \Lbm 
 \ +  \  (\tilde F-Z) \  
\Lap  
 \ + \ 
Z\  \Labr  \Big]\,,
  \\[-4pt]
  \cline{1-2}
\ud Z 
&\ =  \  2(\alpha_1+\alpha_2)\,Z\  \Labg \ ,  
\label{equ:TwoSite-dZ}
\end{align}
where the letters (dlog forms) are
\beq
\begin{aligned}
	\Lap \hskip 3pt\ &\equiv\ \ud\log(X_1+Y)\,, \qquad  \Lam\ \hskip 3pt\equiv\ \ud\log(X_1-Y)\,,\\
	\Lbp \ &\equiv\ \ud\log(X_2+Y)\,, \qquad \hskip 3pt\Lbm\ \equiv\ \ud\log(X_2-Y)\,,\\
	\Labg \ &\equiv\ \ud\log(X_1+X_2)\,.
\end{aligned}
\label{eq:summaryletters}
\eeq
We see that, in addition to the three energy singularities of the flat-space wavefunction, we have two extra singularities with a flipped sign for the internal energy $Y$.

\vskip 4pt
Similarly, the equations for the one-loop bubble in  (\ref{equ:bubble}) (for generic twist parameters) are
\begin{alignat}{5}
    \ud\psi&=\mathrlap{\,\alpha_1\left[\Big(\psi- \sum_i F^{(i)}\Big)\,\includegraphics[valign=c]{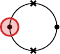}+F^{(\text{I})}\,\includegraphics[valign=c]{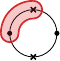}+F^{(\text{II})}\,\includegraphics[valign=c]{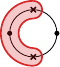}+F^{(\text{III})}\,\includegraphics[valign=c]{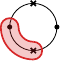}\right]}\nonumber\\&+\mathrlap{ \hspace{0.08333 em}\,\alpha_2\left[\Big(\psi-\sum_i\tilde{F}^{(i)}\Big)\,\includegraphics[valign=c]{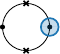}+\tilde{F}^{(\text{I})}\,\includegraphics[valign=c]{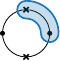}+\tilde{F}^{(\text{II})}\,\includegraphics[valign=c]{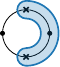}+\tilde{F}^{(\text{III})}\,\includegraphics[valign=c]{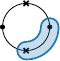}\right],}&\label{equ:BUBBLE1} \\[4pt]
    \cline{1-9}
    \ud F^{(\text{I})}&=\, \alpha_1\,F^{(\text{I})}&&\includegraphics[valign=c]{Bubble_Letters/DlogX1-Y1+Y2.pdf}+\alpha_2\Bigg[\left(\,F^{(\text{I})}\,-\,Z^{(\text{I})}\,\right)&&\includegraphics[valign=c]{Bubble_Letters/DlogX2+Y1+Y2.pdf}+\,Z^{(\text{I})}&&\includegraphics[valign=c]{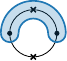}\Bigg], \label{equ:BUBBLE2}\\[4pt]
    \ud F^{(\text{II})}&=\hspace{0.08333 em}\alpha_1 \hspace{0.08333 em}F^{(\text{II})}&&\includegraphics[valign=c]{Bubble_Letters/DlogX1-Y1-Y2.pdf}+\alpha_2\Bigg[\left(\hspace{0.08333 em}F^{(\text{II})}\hspace{0.08333 em}-\hspace{0.08333 em}Z^{(\text{II})}\hspace{0.08333 em}\right)&&\includegraphics[valign=c]{Bubble_Letters/DlogX2+Y1+Y2.pdf}+\,Z^{(\text{II})}&&\includegraphics[valign=c]{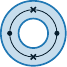}\Bigg],& \label{equ:BUBBLE3}\\[4pt]
    \ud F^{(\text{III})}&=\,\alpha_1 F^{(\text{III})}&&\includegraphics[valign=c]{Bubble_Letters/DlogX1+Y1-Y2.pdf}+\alpha_2\Bigg[\left(F^{(\text{III})}-Z^{(\text{III})}\right)&&\includegraphics[valign=c]{Bubble_Letters/DlogX2+Y1+Y2.pdf}+Z^{(\text{III})}&&\includegraphics[valign=c]{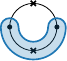}\Bigg],\qquad\qquad\, & \label{equ:BUBBLE4}\\[4pt]
         \cline{1-9}
    \ud Z^{(\text{I})}&=\mathrlap{(\hspace{0.08333 em}\alpha_1\hspace{0.08333 em}+\hspace{0.08333 em}\alpha_2\hspace{0.08333 em})\hspace{0.08333 em}Z^{(\text{I})}\hspace{-0.416667em}\negphantom{Z^{\text{I}}}\hphantom{Z^{\text{II}}}\,\includegraphics[valign=c]{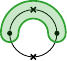}\ , \quad\ \  \ud Z^{(\text{III})}=\,(\alpha_1+\alpha_2)Z^{(\text{III})}\,\includegraphics[valign=c]{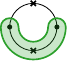}\ , }\label{equ:BUBBLE5}\\[4pt]
     \ud Z^{(\text{II})}&=\mathrlap{\,(\alpha_1+\alpha_2)Z^{(\text{II})}\,\includegraphics[valign=c]{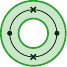}\ , }\label{equ:BUBBLE6}
\end{alignat}
where %$F=F^{(\text{I})}+F^{(\text{II})}+F^{(\text{III})}$ (with an equivalent definition for $\tilde{F}$) and
the letters are 
\beq
\begin{aligned}
    \includegraphics[valign=c]{Bubble_Letters/DlogX1+Y1+Y2.pdf}\hskip 3.5pt&=\ud\log(X_1+Y_1+Y_2)\,,&
    \includegraphics[valign=c]{Bubble_Letters/DlogX2+Y1+Y2.pdf}&=\ud\log(X_2+Y_1+Y_2)\,,\\
    \includegraphics[valign=c]{Bubble_Letters/DlogX1-Y1+Y2.pdf}\hskip 3.5pt&=\ud\log(X_1-Y_1+Y_2)\,,&   \includegraphics[valign=c]{Bubble_Letters/DlogX2-Y1+Y2.pdf}&=\ud\log(X_2-Y_1+Y_2)\,, \\
     \includegraphics[valign=c]{Bubble_Letters/DlogX1-Y1-Y2.pdf}\hskip 3.5pt&=\ud\log(X_1-Y_1-Y_2)\,,
    &  \includegraphics[valign=c]{Bubble_Letters/DlogX2-Y1-Y2.pdf}&=\ud\log(X_2-Y_1-Y_2)\,,
\\
      \includegraphics[valign=c]{Bubble_Letters/DlogX1+Y1-Y2.pdf}\hskip 3.5pt &=\ud\log(X_1+Y_1-Y_2)\,,&
   \includegraphics[valign=c]{Bubble_Letters/DlogX2+Y1-Y2.pdf}&=\ud\log(X_2+Y_1-Y_2)\,,\\
    \includegraphics[valign=c]{Bubble_Letters/DlogX1+X2+2Y1.pdf}&=\ud\log(X_1+X_2+2Y_2)\,,&
    \includegraphics[valign=c]{Bubble_Letters/DlogX1+X2+2Y2.pdf}&=\ud\log(X_1+X_2+2Y_1)\,,&\\
    \includegraphics[valign=c]{Bubble_Letters/DlogX1+X2.pdf}&=\ud\log(X_1+X_2) 
    \,.
\end{aligned}
\label{equ:Bubble-Letters}
\eeq
One might also write down four additional tubes which encircle all the crosses and vertices, but have a break in between one pair of crosses and vertices:
\beq
\includegraphics[valign=c]{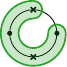}\quad \includegraphics[valign=c]{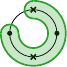}\quad \includegraphics[valign=c]{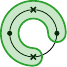}\quad \includegraphics[valign=c]{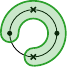}\ .
\label{equ:Bubble-Letters2}
\eeq
However, by converting these tubes into a d$\log$ form, we see that that all represent the same letter, $\ud\log(X_1+X_2)$, since the end of the tube encircling the cross contributes $-Y$, while the other end contributes $+Y$, which cancel. Thus, we have made the choice to include only one representation of this letter, the completely encircled graph in \eqref{equ:Bubble-Letters}. As we will see later, the representations in \eqref{equ:Bubble-Letters2} will not appear in the kinematic flow due to their connection to vanishing canonical forms.\footnote{Note also that unphysical letters such as $X_1+X_2-2Y_1$ and $X_1+X_2-2Y_2$ are automatically excluded by our graphical rule.}

\subsection{Basis Functions}

The starting point of the kinematic flow algorithm is to enumerate the basis functions that will appear in the differential system.
As in~\cite{Arkani-Hamed:2023kig}, we represent each function by a (possibly disconnected) complete tubing of the marked graph (without any nested tubes).  
For the two-site chain, these tubings are  
\begin{equation}\label{equ:2site-fcts}
    \begin{tabular}{c|c|c}
            $\psi$ \includegraphics[valign=c]{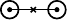} &  $F$ \includegraphics[valign=c]{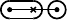}& $Z$ \includegraphics[valign=c]{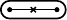}\\
             & $\tilde{F}$ \includegraphics[valign=c]{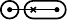}&
    \end{tabular}
\end{equation}
which represent the four functions appearing in (\ref{equ:TwoSite-dPsi})--(\ref{equ:TwoSite-dZ}).
 These tubings naturally arrange themselves into three ``levels" which differ in the number of vertices enclosed in tubes that contain a cross. These levels correspond to the three columns above. Note that the alternating sum of the number of functions in each level vanishes, $1-2+1=0$, which is a general feature of all tree-level graphs.

\vskip 4pt
Similarly, for the one-loop bubble, we have 16 complete tubings of the marked graph
\setlength{\extrarowheight}{10pt}
\begin{equation}\label{equ:bubble-fcts}
    \begin{tabular}{c|cc|ccc}
     $\psi$ \includegraphics[valign=c]{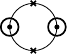} & $F^{(\text{I})}$ \includegraphics[valign=c]{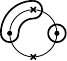}& $\tilde{F}^{(\text{I})}$ \includegraphics[valign=c]{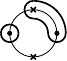}& $Z^{(\text{I})}$ \includegraphics[valign=c]{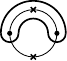}& $\tilde{Z}^{(\text{I})}$ \includegraphics[valign=c]{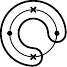}& $G$ \includegraphics[valign=c]{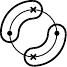}\\
     & $F^{(\text{II})}$ \includegraphics[valign=c]{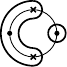}& $\tilde{F}^{(\text{II})}$ \includegraphics[valign=c]{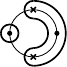}& $Z^{(\text{II})}$ \includegraphics[valign=c]{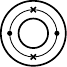} & $\tilde{Z}^{(\text{II})}$ \includegraphics[valign=c]{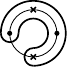} & $\tilde{G}$ \includegraphics[valign=c]{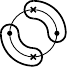}\\ &$F^{(\text{III})}$ \includegraphics[valign=c]{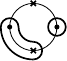}& $\tilde{F}^{(\text{III})}$ \includegraphics[valign=c]{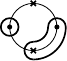} & $Z^{(\text{III})}$ \includegraphics[valign=c]{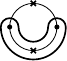} & $\tilde{Z}^{(\text{III})}$ \includegraphics[valign=c]{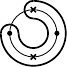}\\&&&&$\tilde{Z}^{(\text{IV})}$ \includegraphics[valign=c]{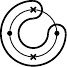}
\end{tabular}
\end{equation}
However, as we see in (\ref{equ:Bubble-Basis}), only $10$ master integrals are needed to fully describe this system. 
As we will explain below, the six tubings denoted by $G$, $\tilde G$, $\tilde Z^{\text{(I-IV)}}$ actually correspond to vanishing functions and therefore don't appear in the differential equations. We will introduce a simple graphical rule to identify graph tubings that should be excluded from the choice of basis functions.

\paragraph{Replacement rules} 

To understand why some of the functions vanish, we first need to explain what the graph tubings in (\ref{equ:2site-fcts}) and (\ref{equ:bubble-fcts}) represent.
First of all, a tubing without any enclosed crosses corresponds to the wavefunction 
\begin{align}
 \Omega_\psi &= \includegraphics[valign=c]{2-site_chain_Forms_and_Letters/twopsi.pdf}  = -\frac{2Y}{B_1 B_2 B_3} \, \ud \omega_1 \wedge \ud \omega_2\,, \\[4pt]
 \Omega_\psi &= \includegraphics[ valign=c]{Bubble_Forms/Psiform.pdf}  = -\frac{4Y_1Y_2}{B_1 B_2B_3}\left(\frac{1}{B_4}+\frac{1}{B_5}\right) \ud\omega_1\wedge\ud\omega_2\,,
  \end{align}
  where the canonical forms were defined in (\ref{equ:Omega-chain}) and (\ref{equ:Omega-bubble}).
 % where $B_1$, $B_2$, $B_3$ were defined in (\ref{Btwositechain}).
 When a tube includes a cross, it indicates that we must replace some of the boundary lines $B_i$ in the canonical form with the twisted lines $T_i = \omega_i$.
 For example, the source functions for the two-site chain are
 \begin{align}
    \Omega_F &=\includegraphics[valign=c]{2-site_chain_Forms_and_Letters/Omega.pdf} =\Omega_{\psi}(B_1\rightarrow T_1)\,, \\
     \Omega_{\tilde F} &=\includegraphics[valign=c]{2-site_chain_Forms_and_Letters/twoFt.pdf} =\Omega_{\psi}(B_2\rightarrow T_2)\,, \\
      \Omega_Z &=\includegraphics[valign=c]{2-site_chain_Forms_and_Letters/twoZ.pdf} =\Omega_{\psi}(B_1\rightarrow T_1, B_2 \to T_2)\,.
\end{align}
We see that we make the replacement $B_i \to T_i$ for any vertex inside a tube that contains a cross. 
Notice that $\Omega_Z$ can be obtained by making an additional replacement from either $\Omega_F$ or $\Omega_{\tilde F}$, and it therefore involves replacing both $B_1$ and $B_2$ from $\Omega_\psi$.  Diagrammatically, this tubing should be understood as the overlap of the two tubings enclosing the central cross:
\begin{equation}
    \includegraphics[valign=c]{2-site_chain_Forms_and_Letters/twoZ.pdf}\equiv\includegraphics[valign=c]{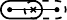}\equiv\includegraphics[valign=c]{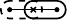}\,.
\end{equation}
Here, the dashed tube represents the pre-existing tube that is being combined with the new solid tube. This is why a tube that encircles multiple vertices with a cross represents multiple plane replacements.

\vskip 4pt
It turns out that all functions appearing in the differential systems for the FRW wavefunction coefficients can be derived through the application of these replacement rules. 
This can be seen to follow from the integration-by-parts structure of the projective simplex forms that define the integrands~\cite{Arkani-Hamed:2023kig}.
The replacement rules are more non-trivial for vertices that have more than one adjacent cross, like in the one-loop bubble. 
Specifically, if a vertex has $m$ adjacent crosses, then there are $2^m-1$ different ways of making replacements. 
For example, the forms $\Omega_{F^{({\rm I},{\rm II},{\rm III})}}$ can be obtained from $\Omega_\psi$ as %\db{explain subtraction; repeat definitions ot $B_{1,4,5}$} %\db{First, all non-vanishing functions}
\begin{align}\label{eq:F2}
    \Omega_{F^{(\text{II})}} &=\includegraphics[valign=c]{Bubble_Forms/F2form.pdf}=\Omega_\psi(B_1,B_4,B_5\rightarrow T_1)\,, \\    \Omega_{F^{(\text{I})}} &=\includegraphics[valign=c]{Bubble_Forms/F1form.pdf}=\Omega_\psi(B_1,B_5\rightarrow T_1)-\Omega_{F^{(\text{II})}}\,,\\
   \Omega_{F^{(\text{III})}}  &=\includegraphics[valign=c]{Bubble_Forms/F3form.pdf}=\Omega_\psi(B_1,B_4\rightarrow T_1)-\Omega_{F^{(\text{II})}}\,.  \label{equ:F3}
\end{align}
As for the two-site chain, we replace all $B_i$ that depend on the energies associated with the crosses with the corresponding twisted lines $T_i$. 
In addition, we subtract any functions associated with larger tubes around that vertex. 
The purpose of this subtraction is to remove all redundant functions to yield a minimal representation of the resulting differential equations.
This procedure straightforwardly generalizes to arbitrarily complex tubings, and the general combinatorial formula can be found in~\cite{Arkani-Hamed:2023kig}.

\paragraph{Vanishing functions} Having established these replacement rules, we can now explain why six of the tubings in \eqref{equ:bubble-fcts} corresponding to vanishing canonical forms.
 They come in two distinct types, $\tilde{Z}$ and $G$, which we will consider separately.
 
 \vskip 4pt
Let us first look at $\tilde{Z}^{(\text{IV})}$ as an explicit example. This canonical form can be obtained from $\Omega_{F^{(\text{II})}}$  by the following replacements
\beq
\label{eq:WZ4}
\begin{aligned}
\Omega_{\tilde{Z}^{(\text{IV})}}= \includegraphics[valign=c]{Bubble_Forms/Z4tildeform.pdf} 
&= \Omega_{F^{(\text{II})}}(B_2,B_4\rightarrow T_2)-\Omega_{F^{(\text{II})}}(B_2,B_4,B_5\rightarrow T_2)\\
&= \Omega_{F^{(\text{II})}}(B_2\rightarrow T_2)-\Omega_{F^{(\text{II})}}(B_2\rightarrow T_2)\\[6pt]
&=0\,.
\end{aligned}
\eeq
In the first equality, the second term is present due to the subtraction rule, explained below~\eqref{equ:F3}. 
In the second equality, we have used that
$\Omega_{F^{(\text{II})}}$ does not contain $B_4$ and $B_5$; cf.~\eqref{eq:F2}. The canonical form $\Omega_{\tilde{Z}^{(\text{IV})}}$ therefore vanishes.
By symmetry, the same argument applies to the other $\Omega_{\tilde Z}$. The four forms $\Omega_{\tilde Z^{\text{(I-IV)}}}$ therefore all vanish and do not have to be included as basis functions.

\vskip 4pt
Similarly,  $G$ and $\tilde{G}$ also vanish. To see this, note that $\Omega_G$ is related to $\Omega_{F^{(\text{I})}}$ by 
\beq
\begin{aligned}
 \Omega_{G}= \includegraphics[valign=c]{Bubble_Forms/Gform.pdf} &= \Omega_{F^{(\text{I})}}(B_2,B_4\rightarrow T_2)-\Omega_{F^{(\text{I})}}(B_2,B_4,B_5\rightarrow T_2)\\
 &=\Omega_{F^{(\text{I})}}(B_2,B_4\rightarrow T_2)-\Omega_{F^{(\text{I})}}(B_2,B_4\rightarrow T_2)\\[6pt]
 &=0\,,
\end{aligned}
\eeq
where we have used that $\Omega_{F^{(\text{I})}}$ doesn't depend on $B_5$; cf.~\eqref{equ:F3}.
By symmetry, $\Omega_{\tilde G}$ also vanishes.

%It may seem at this point that we are done as, we can just follow the same procedure as in flat space and when we write down the equations we must necessarily find that these forms vanish. However, this is not sufficient for two reasons. The first is that it is aesthetically undesirable to have zeros showing up in our equations that we don't know about and so we would like to be able to exclude them immediately. The second is perhaps more important, which is that the kinematic flow presented in \cite{Arkani-Hamed:2023kig} allows $\tilde{Z}^{(\text{IV})}$ to flow into $Z^{(\text{II})}$ and so if we don't have some way to exclude $\tilde{Z}^{(\text{IV})}$ we may end up with erroneous terms in our differential equations. We will therefore need an additional rule to exclude these canonical forms from our basis.

\paragraph{Graphical rule} 

We now introduce a simple graphical rule to identify all vanishing functions directly from the graph tubings.
To state this rule, we dress the tubings with arrows pointing from each cross inside a tube to the neighboring tube. This is best illustrated by an example. 
For the tubings in (\ref{equ:bubble-fcts}), we have  %\db{Put a vertical line between the first 4 columns and the last 2 to separate the vanishing functions} %\db{arrows in red, so that they are easier to see}
\setlength{\extrarowheight}{10pt}
\begin{equation}\label{tab:arrowtubes}
    \begin{tabular}{c|cc|c||cc}
     $\psi$ \includegraphics[valign=c]{Bubble_Forms/Psiform.pdf} & $F^{(\text{I})}$ \includegraphics[valign=c]{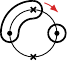}& $\tilde{F}^{(\text{I})}$ \includegraphics[valign=c]{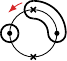}& $Z^{(\text{I})}$ \includegraphics[valign=c]{Bubble_Forms/Z1form.pdf}& $\tilde{Z}^{(\text{I})}$ \includegraphics[valign=c]{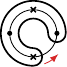}& $G$ \includegraphics[valign=c]{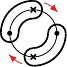}\\
     & $F^{(\text{II})}$ \includegraphics[valign=c]{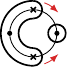}& $\tilde{F}^{(\text{II})}$ \includegraphics[valign=c]{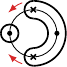}& $Z^{(\text{II})}$ \includegraphics[valign=c]{Bubble_Forms/Z2form.pdf} & $\tilde{Z}^{(\text{II})}$ \includegraphics[valign=c]{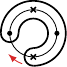} & $\tilde{G}$ \includegraphics[valign=c]{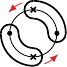}\\ &$F^{(\text{III})}$ \includegraphics[valign=c]{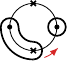}& $\tilde{F}^{(\text{III})}$ \includegraphics[valign=c]{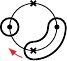}& $Z^{(\text{III})}$ \includegraphics[valign=c]{Bubble_Forms/Z3form.pdf} & $\tilde{Z}^{(\text{III})}$ \includegraphics[valign=c]{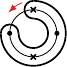}\\&&&&$\tilde{Z}^{(\text{IV})}$\includegraphics[valign=c]{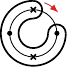}
\end{tabular}
\end{equation}
We can then state the rule which determines if a tubing corresponds to a vanishing function: 
\begin{center}
{\it A function vanishes if we can start from a tube and return to it by following the arrows.}
\end{center}
Applying this rule to the tubings in \eqref{tab:arrowtubes}, we see that the six functions $G$, $\tilde G$, $\tilde Z^{(\rm I-IV)}$ vanish. We have therefore correctly recovered the 10 basis functions given in (\ref{equ:Bubble-Basis}).  As we will see below, the same rule will work in more nontrivial examples. We also note that the only tubings that can describe a vanishing function are those in the last level, with all vertices enclosed in tubes with crosses. This is a general feature that will hold in all examples.

\paragraph{Counting functions} Without taking the vanishing of functions into account, the tubing prescription produces $4^E$ functions (where $E$ is the number of edges), just as in the tree-level case.
However, to determine the actual size of the basis, we must subtract the number of vanishing functions. This is easy to do in specific examples. Let us illustrate this for the case of the $n$-gon loop (a loop with $n$ sites and $n$ edges).

\vskip 4pt
For the one-loop bubble, we had two types of  vanishing functions:
\begin{equation}
    \includegraphics[valign=c]{Bubble_Forms/Z1tildearrow.pdf}\quad\includegraphics[valign=c]{Bubble_Forms/Garrow.pdf}\ ,
\end{equation}
which have one or two arrows, respectively.
Similarly, for the one-loop triangle, we have three types of vanishing functions: 
\begin{align}
    \includegraphics[valign=c]{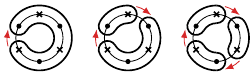}\ ,
\end{align}
with one, two and three arrows, respectively. This behaviour is generic: For an $n$-gon loop, there are vanishing tubings corresponding to placing $i\in(1,n)$ arrows on the edges of the loop. The total number of each of these types of tubings is then equal to the number of ways to choose $i$ out of $n$ edges. As the arrows can go either way around the loop we must double this counting. The total number of vanishing function therefore is
\begin{align}
  2  \sum_{i=1}^n \binom{n}{i}=2\hs (2^n-1)\,,
  \label{equ:count}
\end{align}
and the total number of non-vanishing functions becomes
\begin{align}\label{eq:Ngoncount}
   N_{\rm basis}^{\text{$n$-gon}} = 4^n-2(2^n-1)\,,
\end{align}
which is in agreement with the counting presented in~\cite{He:2024olr}. 
For the one-loop bubble, the counting in   \eqref{equ:count} gives $4 + 2 =6$ vanishing functions, which corresponds to the functions  $\{\tilde Z^{(\rm I-IV)}\}$ and $\{G,\tilde G\}$ in \eqref{tab:arrowtubes}, respectively. The basis then has $16-6=10$ non-vanishing functions. Similarly, for the one-loop triangle, we have a total of $6 + 6 + 2 = 14$ vanishing functions and $64-14=50$ non-vanishing basis functions.
 
\vskip 4pt
It is also interesting to count the number of functions at each level. For the $n$-gon loop, the total number of tubings at level $l$ is
\begin{align}
    N_l^{\text{$n$-gon}}=3^{l}\binom{n}{l}\,,
\end{align}
where the counting starts at level $l=0$ for the wavefunction.
The final level is $l =n$ from which we must remove the $2(2^n-1)$ vanishing forms. At tree level, it was observed in \cite{Arkani-Hamed:2023kig} that the alternating sum of the number of functions at each level vanishes,
\begin{align}
    \sum_{l=0}^{n}(-1)^{l} N^{\text{tree}}_l=0\,.
\end{align}
However, this is no longer true at one loop. Instead, we find that this alternating sum of all functions (including vanishing functions) gives
\begin{align}\label{eq:altsum}
    \sum_{l=0}^{n}(-1)^{l}   N_l^{\text{$n$-gon}} 
    =(-2)^n\,.
\end{align}
Removing the vanishing functions from the final level, the right-hand side would instead become $(-1)^{n}(2-2^n)$. 
For the one-loop bubble and triangle, the alternating sums (taking into account the vanishing functions) are $1-6+3=-2$ and $1-9+27-13 = 6$, respectively. It would be nice to understand the systematics of these alternating sums for arbitrary loop graphs.

\subsection{Rules of the Kinematic Flow}
\label{ssec:Rules}

In~\cite{Arkani-Hamed:2023kig}, simple and universal graphical rules were discovered that predict the differential equations for arbitrary tree graphs. Here, we show that the same rules also determine the equations for loop integrands if we account for the new possibility that some basis functions can vanish.

\vskip 4pt
The rules of the kinematic flow are easiest to state with a specific example. 
We start with the graph tubing associated to a ``parent function" of interest:
\begin{equation}
    \includegraphics{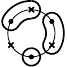}
     \nonumber
\end{equation}
Next, we generate a family tree of its ``descendants":
\begin{enumerate}
    \item {\it Activation}. We first activate each 
    of the disjoint tubes:
    \begin{equation}
        \includegraphics{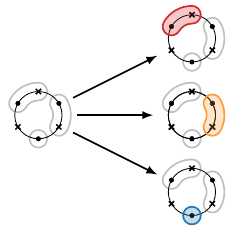}
         \nonumber
    \end{equation}
    The colored (shaded) tubes will become letters in the differential equation.
    \item {\it Growth and Merger}. An activated tube without a cross can {\it grow} to incorporate any adjacent crosses. If this causes two tubes to overlap, they {\it merge}. For our example, this leads to
    \begin{equation}
        \includegraphics{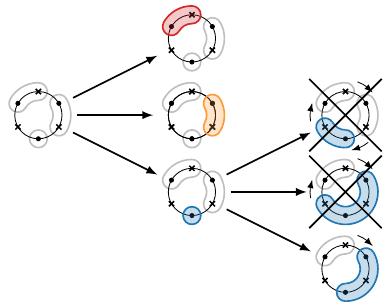}
         \nonumber
    \end{equation}  
    In the first two branches, the activated tubes contain a cross and therefore don't grow. 
    In the bottom branch, the activated tube grows to enclose the neighboring crosses. In principle, this can happen in three different ways, but two of these lead to vanishing functions.
   
    \item {\it Absorption}. If an activated tube has an arrow that points towards any tubes containing a cross, then it absorbs these tubes. For our example, we have 
    \begin{equation}        \includegraphics{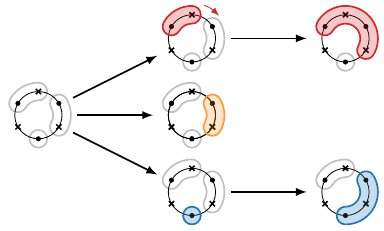}
         \nonumber
    \end{equation}  
    We see one such absorption in the top branch. 
\end{enumerate}
Having established the kinematic flow for a function, we can write down the corresponding differential equation by applying the same algorithm as in the tree-level case~\cite{Arkani-Hamed:2023kig}:
\begin{enumerate}
    \item Assign to each graph tubing its corresponding function multiplied by $(-1)^{N_a}$, where $N_a$ is the number of absorptions required to reach that point on the family tree.
    \item Assign a letter to each graph tubing based on the activated tube.
    \item Multiply each letter by the function associated to the graph tubing {\it minus} its immediate descendant(s). Include an overall  factor of $\sum_v \alpha_v$, where $v$ are the vertices inside the tube that was activated in the first step.
\end{enumerate}
Equipped with these rules, we can write down the differential equation for the chosen parent function from our kinematic flow diagram:
\begin{equation}
    \begin{aligned}
        \ud \ \includegraphics[valign=c]{Triangle_Letters_and_Forms/F110.pdf}&=\alpha_1\left(\includegraphics[valign=c]{Triangle_Letters_and_Forms/F110.pdf}+\includegraphics[valign=c]{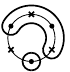}\right)\includegraphics[valign=c]{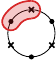}-\alpha_1\includegraphics[valign=c]{Triangle_Letters_and_Forms/F120.pdf}\includegraphics[valign=c]{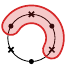}+\alpha_2\includegraphics[valign=c]{Triangle_Letters_and_Forms/F110.pdf}\ \includegraphics[valign=c]{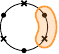}\\&+\alpha_3\left(\includegraphics[valign=c]{Triangle_Letters_and_Forms/F110.pdf}\hskip 1pt -\hskip 1pt \includegraphics[valign=c]{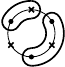}\hskip 1pt \right)\hskip 4pt \includegraphics[valign=c]{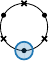}+\alpha_3\includegraphics[valign=c]{Triangle_Letters_and_Forms/F113.pdf}\ \hskip 2pt  \includegraphics[valign=c]{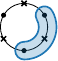}\ .
    \end{aligned}
\end{equation}
This is just a single element of the full set of differential equations required to calculate the wavefunction coefficient of the one-loop triangle. However, the procedure can easily be applied to all basis functions to close the system of differential equations (see Section~\ref{ssec:triangle}).

%\newpage
\subsection{Application to Selected Loops}

In this section, we will apply the kinematic flow rules to a few selected examples. We will present the differential equations produced by the kinematic flow, which we checked against explicit computations following the same approach as in \cite{Arkani-Hamed:2023kig}. 

\subsubsection{One-Loop Bubble}

In the following, we will re-derive equations (\ref{equ:BUBBLE1}) -- (\ref{equ:BUBBLE6}) for the one-loop bubble from the flow rules. In \eqref{tab:arrowtubes}, we tabulated all non-vanishing basis functions. We start by constructing the evolutionary trees for each of these functions.

\begin{itemize}
\item {\bf Level 1:} The tree for the wavefunction is
\begin{equation}
    \psi:\ \includegraphics[valign=c]{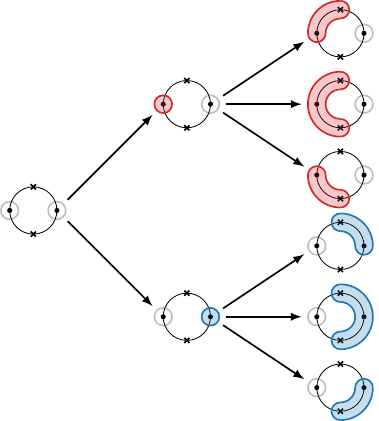}
    \label{equ:PhiFlow2}
\end{equation}
The tubes around each vertex are activated in the first step and then grow to encircle the neighboring crosses in the second step. Assigning the corresponding letters and functions to each tubing in the tree, we obtain (\ref{equ:BUBBLE1}).

\item {\bf Level 2:} Next, we consider the source function $F^{({\rm I})}$. Its tree is 
\begin{equation}
    F^{(\text{I})}:\ \includegraphics[valign=c]{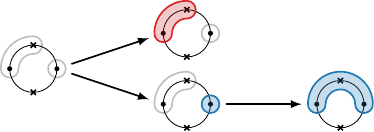}
    \label{equ:F1Flow}
\end{equation}
Note that in lower branch only one new function is generated through growth and merger, although the tube can grow in three different ways. The other two tubings correspond to the vanishing functions $G$ and $\tilde Z^{({\rm II})}$. 
Equation (\ref{equ:BUBBLE2}) then immediately follows. By symmetry, the equations for $F^{({\rm III})}$, $\tilde F^{({\rm I})}$, $\tilde F^{({\rm III})}$ are derived in the same way.

\vskip 4pt
For the function $F^{({\rm II})}$, we have the following tree
\begin{equation}
    F^{(\text{II})}:\ \includegraphics[valign=c]{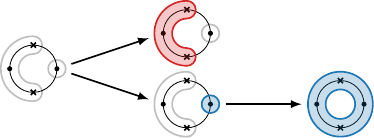}
\end{equation}
\\
which gives equation (\ref{equ:BUBBLE3}). As in (\ref{equ:F1Flow}), we generate only one non-vanishing function in the final step. By symmetry, the equation for $\tilde F^{({\rm II})}$ is derived in the same way.

%\db{comment on vanishing functions}

\item {\bf Level 3:} Finally, for the functions $Z^{({\rm I,II})}$, we only have activation: 
\begin{equation}
    Z^{(\text{I})}:\ \includegraphics[valign=c]{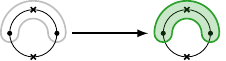}
\end{equation}
\begin{equation}
    Z^{(\text{II})}:\ \includegraphics[valign=c]{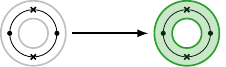}
\end{equation}
By symmetry, the function $Z^{({\rm III})}$ has the same tree as the function $Z^{({\rm I})}$. This leads directly to equations (\ref{equ:BUBBLE5}) and (\ref{equ:BUBBLE6}).
\end{itemize}
We see that the equations for the one-loop bubble are as easy to derive as those for the two-site chain if we take into account that some functions in the evolutionary tree can vanish.

\subsubsection{One-Loop Triangle}
\label{ssec:triangle}

A more non-trivial example is the one-loop triangle. We again start from the flat-space wavefunction which can be derived as the sum of six compatible graph tubings:
\begin{align}
\hat{\psi}^{\text{flat}}_{(3),\text{triangle}} &=\includegraphics[valign=c]{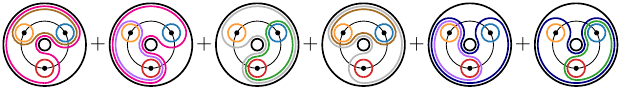} \nonumber \\[4pt] 
    &=-\frac{8Y_1Y_2Y_3}{\color{newblue}\hat B_1\color{neworange}\hat B_2\color{newred}\hat B_3\color{black} \hat B_{10}}\left(\frac{1}{\color{brown}\hat B_4\color{magenta}\hat B_7}+\frac{1}{\color{newpurple}\hat B_5\color{magenta}\hat B_7}+\frac{1}{\color{newgreen}\hat B_6\color{grey}\hat B_8}+\frac{1}{\color{brown}\hat B_4\color{grey}\hat B_8}+\frac{1}{\color{newpurple}\hat B_5\color{navy}\hat B_{9}}+\frac{1}{\color{newgreen}\hat B_6\color{navy}\hat B_9}\right) ,
\end{align}
where
\beq\label{eq:BTriangle}
\begin{aligned}
    \hat B_1&=X_1+Y_1+Y_3\,,&\hat B_{2}&=X_2+Y_1+Y_2\,,&\hat B_3&=X_3+Y_2+Y_3\,,\\
    \hat B_4&=X_1+X_2+Y_2+Y_3\,,&\hat B_5&=X_1+X_3+Y_1+Y_2\,,&\hat B_6&=X_2+X_3+Y_1+Y_3\,,\\
    \hat B_7&=X_1+X_2+X_3+2Y_2\,,& \hat B_8&=X_1+X_2+X_3+2Y_3\,,& \hat B_9&=X_1+X_2+X_3+2Y_1\,,\\
    \hat B_{10}&=X_1+X_2+X_3\,.
\end{aligned}
\eeq
Substituting this into \eqref{eq:Loop} gives the relevant integral for the cosmological wavefunction.

\paragraph{Basis functions} The following complete tubings of the marked graph define the non-vanishing basis functions: 
\begin{equation}\label{eq:TriangleForms}
    \begin{tabular}{ccccc}
        $F_{000}$ \includegraphics[valign=c,padding = 0ex 0.5ex 0ex 0.5ex]{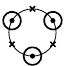} \\\hline 
        $F_{100}$ \includegraphics[valign=c]{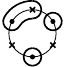}&$F_{200}$ \includegraphics[valign=c,padding = 0ex 0.5ex 0ex 0.5ex]{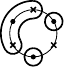}&$F_{300}$ \includegraphics[valign=c,padding = 0ex 0.5ex 0ex 0.5ex]{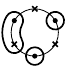}\\\hline
        $F_{110}$ \includegraphics[valign=c,padding = 0ex 0ex 0ex 0.5ex]{Triangle_Letters_and_Forms/F110.pdf}&$F_{120}$ \includegraphics[valign=c]{Triangle_Letters_and_Forms/F120.pdf}&$F_{130}$ \includegraphics[valign=c,padding = 0ex 0ex 0ex 0.5ex]{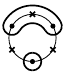}&$F_{210}$ \includegraphics[valign=c]{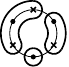}&$F_{220}$ \includegraphics[valign=c,padding = 0ex 0ex 0ex 0.5ex]{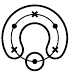}\\
        $F_{330}$ \includegraphics[valign=c,padding = 0ex 0.5ex 0ex 0ex]{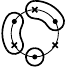}&$F_{230}$ \includegraphics[valign=c]{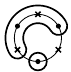}&$F_{310}$ \includegraphics[valign=c,padding = 0ex 0.5ex 0ex 0ex]{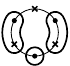}&$F_{320}$ \includegraphics[valign=c,padding = 0ex 0.5ex 0ex 0ex]{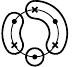}\\\hline
        $F_{113}$ \includegraphics[valign=c,padding = 0ex 0ex 0ex 0.5ex]{Triangle_Letters_and_Forms/F113.pdf}&$F_{123}$ \includegraphics[valign=c]{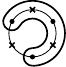}&$F_{132}$ \includegraphics[valign=c,padding = 0ex 0ex 0ex 0.5ex]{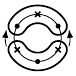}&$F_{133}$ \includegraphics[valign=c,padding = 0ex 0ex 0ex 0.5ex]{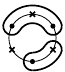}\\
        $F_{131}$ \includegraphics[valign=c]{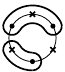}&$F_{231}$ \includegraphics[valign=c]{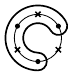}&$F_{213}$ \includegraphics[valign=c]{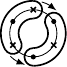}&$F_{331}$ \includegraphics[valign=c]{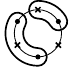}&$F_{222}$ \includegraphics[valign=c]{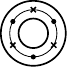}\\
        $F_{311}$ \includegraphics[valign=c]{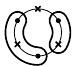}&$F_{312}$ \includegraphics[valign=c]{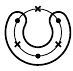}&$F_{321}$ \includegraphics[valign=c]{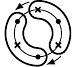}&$F_{313}$ \includegraphics[valign=c]{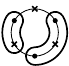}
    \end{tabular}
\end{equation}
% \begin{equation}\label{eq:TriangleForms}
%     \includegraphics[valign=c]{Triangle_Letters_and_Forms/TriangleForms.pdf}
% \end{equation}
For each vertex, there are three different ways to encircle the neighboring crosses. The ordered subscripts on
 $F_{ijk}$ denote the different ways in which these elementary tubings can be combined. 
In total, there are 50 non-vanishing functions: the 26 shown in \eqref{eq:TriangleForms} plus 24 similar tubings formed by permuting the zeros in $F_{i00}$ and $F_{ij0}$. 
%These 50 functions come in 13 types which are 
%\begin{align}\nonumber
 %   \psi&=F_{000}\,,& F_1&=F_{100}\,,& Q_1&=F_{200}\,,&q_3^{(1)}&=F_{110}\,,& q_2^{(1)}&=F_{120}\,, & h_1&=F_{210}\,, &q_1^{(1)}&=F_{130}\,,\\ \tilde{f}_1&=F_{220}\,,&f_1&=F_{310}\,,& G_2&=F_{113}\,,&Z_3&=F_{123}\,,&\tilde{Z}_1&=F_{213}\,,& Z_0&=F_{222}\,.
%\end{align}
The differential equations for each of these basis functions can then be deduced from our kinematic flow rules.

\paragraph{Level 1:} The tree for the wavefunction is
\begin{equation}
    F_{000}:\includegraphics[valign=c]{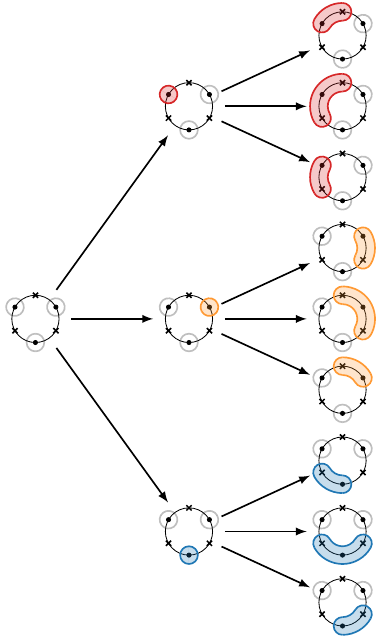}
\end{equation}
and the associated  differential equation is 
\begin{align}\nonumber
    \ud F_{000}&=\alpha_1\Bigg[\left(F_{000}-F_{100}-F_{200}-F_{300}\right)\,\includegraphics[valign=c]{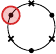}+F_{100}\,\includegraphics[valign=c]{Triangle_Letters_and_Forms/dlogB1-+.pdf}+F_{200}\,\includegraphics[valign=c]{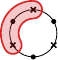}+F_{300}\,\includegraphics[valign=c]{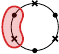}\hskip 1pt\Bigg]\\&+
    \alpha_2\Bigg[\left(F_{000}-F_{010}-F_{020}-F_{030}\right)\,\includegraphics[valign=c]{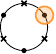}+F_{010}\,\includegraphics[valign=c]{Triangle_Letters_and_Forms/dlogB2+-.pdf}+F_{020}\,\includegraphics[valign=c]{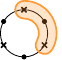}+F_{030}\,\includegraphics[valign=c]{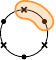}\Bigg]\\\nonumber&+
    \alpha_3\Bigg[\left(F_{000}-F_{001}-F_{002}-F_{003}\right)\,\includegraphics[valign=c]{Triangle_Letters_and_Forms/dlogB3++.pdf}\hskip 3pt+F_{001}\,\includegraphics[valign=c]{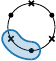}+F_{002}\,\includegraphics[valign=c]{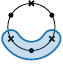}+F_{003}\,\includegraphics[valign=c]{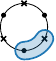}\hskip 1.5pt\Bigg]\,.
\end{align}
This equation is analogous to  \eqref{equ:PhiFlow2} for the one-loop bubble.

\paragraph{Level 2:} We have found $9$ new basis functions, but they come in just two distinct forms, which can be illustrated by the functions $F_{100}$ and $F_{200}$. The flow for the function $F_{100}$ is
\begin{equation}
    F_{100}:\raisebox{-0.5\height}{\includegraphics[valign=c]{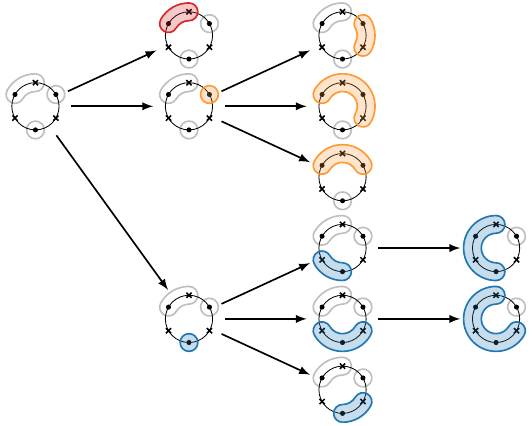}}
\end{equation}
We see first activation, then growth and merger (in the middle branch), and finally absorption (in the bottom branch). The relevant differential equation then is
\begin{align}\nonumber
        \ud F_{100}&=\alpha_1F_{100}\,\includegraphics[valign=c]{Triangle_Letters_and_Forms/dlogB1-+.pdf}\\\nonumber &+ \alpha_2\left[\left(F_{100}-F_{110}-F_{120}-F_{130}\right)\,\includegraphics[valign=c]{Triangle_Letters_and_Forms/dlogB2++.pdf}\hskip 1pt+\hskip 1ptF_{110}\hskip 3pt\,\includegraphics[valign=c]{Triangle_Letters_and_Forms/dlogB2+-.pdf}\hskip 1pt+\hskip 2ptF_{120}\,\includegraphics[valign=c]{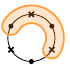}\hskip 1pt+\hskip 1ptF_{130}\,\includegraphics[valign=c]{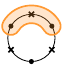}\right]\\\nonumber&+ \alpha_3\left[\left(F_{100}- F_{101}-F_{102}-F_{103}\right)\,\includegraphics[valign=c]{Triangle_Letters_and_Forms/dlogB3++.pdf}+(F_{101}+F_{201})\,\includegraphics[valign=c]{Triangle_Letters_and_Forms/dlogB3+-.pdf}+(F_{102}+F_{202})\,\includegraphics[valign=c]{Triangle_Letters_and_Forms/dlogB3--.pdf}\right.\\&\left. \hskip 25pt+F_{103}\,\includegraphics[valign=c]{Triangle_Letters_and_Forms/dlogB3-+.pdf}-F_{201}\,\includegraphics[valign=c]{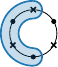}-F_{202}\,\includegraphics[valign=c]{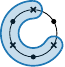}\right].
\end{align}
The second type of function can be illustrated by  $F_{200}$, whose flow is
\begin{equation}
    F_{200}:\raisebox{-0.5\height}{\includegraphics[valign=c]{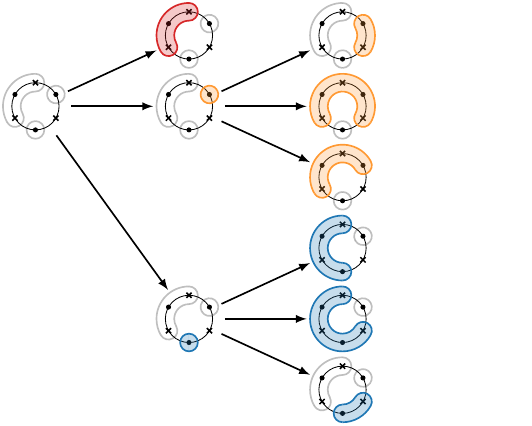}}
\end{equation}
We see the standard activation (in step 1), as well as growth and merger (in step 2). The differential equation associated to this tree structure is
\begin{align}
        \ud F_{200}&=\alpha_1F_{200}\,\includegraphics[valign=c]{Triangle_Letters_and_Forms/dlogB1--.pdf}\\ &+\alpha_2\left[\left(F_{200}-F_{210}-F_{220}-F_{230}\right)\,\includegraphics[valign=c]{Triangle_Letters_and_Forms/dlogB2++.pdf}+F_{210}\hskip 4.5pt\,\includegraphics[valign=c]{Triangle_Letters_and_Forms/dlogB2+-.pdf}+F_{220}\,\includegraphics[valign=c]{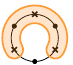}+F_{230}\,\includegraphics[valign=c]{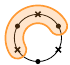}\right]\nn\\&+\alpha_3\left[\left(F_{200}-F_{201}-F_{202}-F_{203}\right)\,\includegraphics[valign=c]{Triangle_Letters_and_Forms/dlogB3++.pdf}\hskip 4pt+F_{201}\,\includegraphics[valign=c]{Triangle_Letters_and_Forms/dlogB5-+.pdf}\hskip 4pt+F_{202}\hskip 1.5pt\,\includegraphics[valign=c]{Triangle_Letters_and_Forms/dlogB5--.pdf}\hskip 3pt+F_{203}\hskip 6.5pt\,\includegraphics[valign=c]{Triangle_Letters_and_Forms/dlogB3-+.pdf}\right].\nn
\end{align}
Unlike the case of the one-loop bubble, we don't have any vanishing functions at this stage.

\paragraph{Level 3:} The previous level has produced 6 distinct source functions, whose kinematic flow we will now present. 
\begin{itemize}
\item The tree for the function $F_{110}$ is
\begin{equation}
    F_{110}:\includegraphics[valign=c]{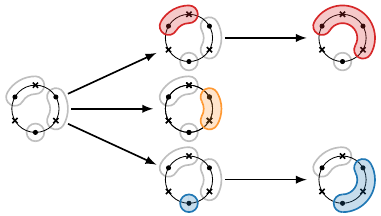}
\end{equation}
In the top branch, we have first activation and then absorption; in the middle branch, we have only activation; and, in the bottom branch, we have activation, growth and merger. We only have one type of growth/merger, because the others lead to vanishing functions. The differential equation predicted by the above tree is
\begin{align}
    \begin{aligned}
        \ud F_{110}&=\alpha_1\Bigg[(F_{110}+F_{120})\,\includegraphics[valign=c]{Triangle_Letters_and_Forms/dlogB1-+.pdf}-F_{120}\,\includegraphics[valign=c]{Triangle_Letters_and_Forms/dlogB4-+red.pdf}\Bigg]+\alpha_2F_{110}\,\includegraphics[valign=c]{Triangle_Letters_and_Forms/dlogB2+-.pdf}\\&+\alpha_3\hskip 1pt\Bigg[(F_{110}-F_{113})\hskip 4pt\,\includegraphics[valign=c]{Triangle_Letters_and_Forms/dlogB3++.pdf}+F_{113}\hskip 3.5pt\,\includegraphics[valign=c]{Triangle_Letters_and_Forms/dlogB6++.pdf}\hskip 2.5pt\Bigg].
    \end{aligned}
\end{align}
\vskip 4pt
\item The tree for the function $F_{120}$ is
\begin{equation}
    F_{120}:\includegraphics[valign=c]{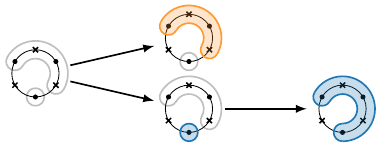}
    \label{equ:dF120}
\end{equation}
In the bottom branch, we only get one merger, since the other two possible mergers lead to vanishing functions. 
Something similar happens for $F_{220}$:
\begin{equation}
    F_{220}:\includegraphics[valign=c]{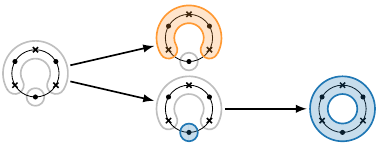}
    \label{equ:dF220}
\end{equation}
The differential equations corresponding to (\ref{equ:dF120}) and (\ref{equ:dF220}) are
\begin{align}
    \ud F_{120}&=(\alpha_1+\alpha_2)F_{120}\,\includegraphics[valign=c]{Triangle_Letters_and_Forms/dlogB4-+.pdf}\hskip 2pt+\alpha_3\left[(F_{120}-F_{123})\,\includegraphics[valign=c]{Triangle_Letters_and_Forms/dlogB3++.pdf}+F_{123}\,\includegraphics[valign=c]{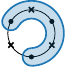}\right] , \\
      \ud F_{220}&=(\alpha_1+\alpha_2)F_{220}\,\includegraphics[valign=c]{Triangle_Letters_and_Forms/dlogB4--.pdf}+\alpha_3\left[ (F_{220}-F_{222})\,\includegraphics[valign=c]{Triangle_Letters_and_Forms/dlogB3++.pdf}+F_{222}\,\includegraphics[valign=c]{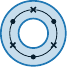}\right] .
\end{align}
\item The tree for the function $F_{210}$ is
\begin{equation}
    F_{210}:\includegraphics[valign=c]{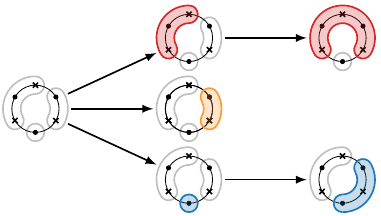}
\end{equation}
Vanishing functions again play an important role in this tree structure. In top branch, we only get one absorption, since the other two possibilities give vanishing functions. The same applies to the merger in the bottom branch. The associated differential equation then is
\begin{align}
    \begin{aligned}
        \ud F_{210}&=\alpha_1\Bigg[(F_{200}+F_{220})\,\includegraphics[valign=c]{Triangle_Letters_and_Forms/dlogB1--.pdf}-F_{220}\,\includegraphics[valign=c]{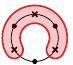}\Bigg]+\alpha_2F_{210}\, \includegraphics[valign=c]{Triangle_Letters_and_Forms/dlogB2+-.pdf}\\&+\alpha_3\hskip 1pt\Bigg[(F_{210}-F_{213})\hskip 4pt\,\includegraphics[valign=c]{Triangle_Letters_and_Forms/dlogB3++.pdf}+F_{213}\hskip 5.5pt\,\includegraphics[valign=c]{Triangle_Letters_and_Forms/dlogB6++.pdf}\hskip 2.5pt\Bigg]\,.
    \end{aligned}
\end{align}
\item The tree for the function $F_{130}$ is
\begin{equation}
    F_{130}:\raisebox{-\height/3}{\includegraphics[valign=c]{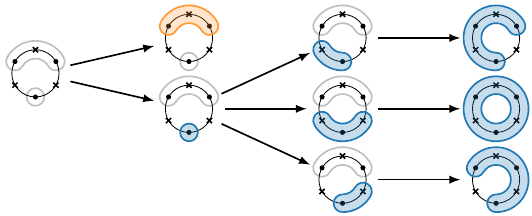}}
\end{equation}
which contains the standard activation, growth and absorption. In the final step, we have excluded some vanishing functions.

The corresponding differential equation is
\begin{align}
        \ud F_{130} &=(\alpha_1+\alpha_2)F_{130}\,\includegraphics[valign=c]{Triangle_Letters_and_Forms/dlogB4++.pdf}  \\
        &+\alpha_3\Bigg[(F_{130}-F_{131}-F_{132}-F_{133})\,\includegraphics[valign=c]{Triangle_Letters_and_Forms/dlogB3++.pdf}\hskip 4pt+(F_{131}+F_{231})\,\includegraphics[valign=c]{Triangle_Letters_and_Forms/dlogB3+-.pdf} -F_{231}\,\includegraphics[valign=c]{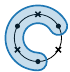} \nn\\& \hskip 15pt+ (F_{132}+F_{222})\,\includegraphics[valign=c]{Triangle_Letters_and_Forms/dlogB3--.pdf}-F_{222}\, \includegraphics[valign=c]{Triangle_Letters_and_Forms/dlogB10.pdf}+(F_{133}+F_{123})\hskip 3pt\,\includegraphics[valign=c]{Triangle_Letters_and_Forms/dlogB3-+.pdf}-F_{123}\,\includegraphics[valign=c]{Triangle_Letters_and_Forms/dlogB8.pdf}\Bigg]\,.\nn
\end{align}
Note that the activated letter in the first line contains two vertices, so that overall factor is $\alpha_1+\alpha_2$.
\item The tree for $F_{310}$ is
\begin{equation}
    F_{310}:\includegraphics[valign=c]{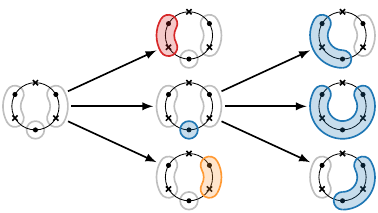}
\end{equation}
which is explained by the standard activation, growth and merger. The corresponding differential equation is
\begin{align}
        \ud F_{310}&=\alpha_1F_{310}\,\includegraphics[valign=c]{Triangle_Letters_and_Forms/dlogB1+-.pdf}+\alpha_2F_{310}\, \includegraphics[valign=c]{Triangle_Letters_and_Forms/dlogB2+-.pdf}\\&+\alpha_3\Bigg[(F_{310}-F_{311}-F_{312}-F_{313})\,\includegraphics[valign=c]{Triangle_Letters_and_Forms/dlogB3++.pdf}+F_{311}\,\includegraphics[valign=c]{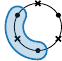}+F_{312}\,\includegraphics[valign=c]{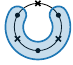}+F_{313}\,\includegraphics[valign=c]{Triangle_Letters_and_Forms/dlogB6++.pdf}\Bigg]\,.\nn
\end{align}
\end{itemize}

\paragraph{Level 4:} The final level contains just 4 distinct functions. 
\begin{itemize}
\item The first two can grow by absorption
\begin{equation}
    F_{113}:\includegraphics[valign=c]{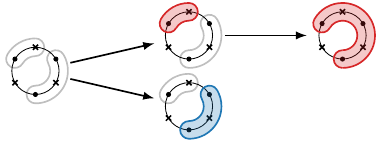}
\end{equation}
\begin{equation}
    F_{213}:\includegraphics[valign=c]{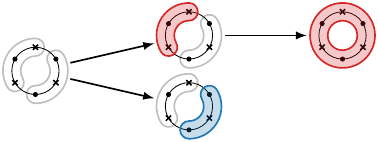}
\end{equation}
so their differential equations are
\begin{align}
    \begin{aligned}
        \ud F_{113}&=\alpha_1\left[(F_{113}+F_{123})\,\includegraphics[valign=c]{Triangle_Letters_and_Forms/dlogB1-+.pdf}-F_{123}\hskip 2pt\,\includegraphics[valign=c]{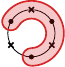}\right]+(\alpha_2+\alpha_3)F_{113} \,\includegraphics[valign=c]{Triangle_Letters_and_Forms/dlogB6++.pdf}\,,\\\ud F_{213}&=\alpha_1\left[(F_{213}+F_{222})\,\includegraphics[valign=c]{Triangle_Letters_and_Forms/dlogB1--.pdf}-F_{222}\,\includegraphics[valign=c]{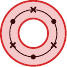}\right]+(\alpha_2+\alpha_3)F_{213}\, \includegraphics[valign=c]{Triangle_Letters_and_Forms/dlogB6++.pdf}\,.
    \end{aligned}
\end{align}
\item The remaining two functions cannot grow in any way, so they just return themselves when we take their derivatives
\begin{align}
    \ud F_{123}&=(\alpha_1+\alpha_2+\alpha_3)F_{123}\,\includegraphics[valign=c]{Triangle_Letters_and_Forms/dlogB8.pdf}\,,\\
    \ud F_{222}&=(\alpha_1+\alpha_2+\alpha_3)F_{222} \,\includegraphics[valign=c]{Triangle_Letters_and_Forms/dlogB10.pdf}\,.
\end{align}
\end{itemize}
This completes our predictions for the differential equations of the one-loop triangle.
%Therefore, we can construct the entire differential equation for $\psi$ by systematic application of the rules that we laid out in the previous section. 

\subsubsection{One-Loop Frying Pan}

Next, we consider the ``one-loop frying pan":
\beq
\includegraphics[valign=c]{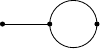}
\eeq
It is straightforward to construct the flat-space wavefunction associated to this graph following the algorithm in Section~\ref{ssec:tubings}.

\vskip 4pt
The basis functions of the associated integral family are 
\begin{equation}
\hspace{-4pt}
\begin{tabular}{l|lll}
    $\psi\negphantom{\psi}\hphantom{F}$ \includegraphics[valign=c,padding = 0ex 0.5ex 0ex 0.5ex]{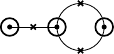}  &
    \\\hline
    $F$ \includegraphics[valign=c,padding = 0ex 0.5ex 0ex 0.5ex]{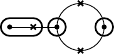}  $\, Q_1$ \includegraphics[valign=c,padding = 0ex 0.5ex 0ex 0.5ex]{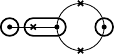}& $Q_2^{(\text{I})}$ \includegraphics[valign=c,padding = 0ex 0.5ex 0ex 0.5ex]{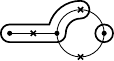} & 
    $Q_3^{(\text{I})}\negphantom{Q_3^{(\text{I})}}\hphantom{Q_2^{(\text{I})}}$ \includegraphics[valign=c,padding = 0ex 0.5ex 0ex 0.5ex]{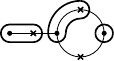}&$\tilde{F}^{(\text{I})}\negphantom{\tilde{F}^{(\text{I})}}\hphantom{Q_2^{(\text{I})}}$ \includegraphics[valign=c,padding = 0ex 0.5ex 0ex 0.5ex]{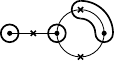}\\\hline
     $q_1\negphantom{q_1}\hphantom{F}$ \includegraphics[valign=c,padding = 0ex 0.5ex 0ex 0.5ex]{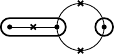}
     &$f^{(\text{I})}\negphantom{f^{(\text{I})}}\hphantom{Q_2^{(\text{I})}}$ \includegraphics[valign=c,padding = 0ex 0.5ex 0ex 0.5ex]{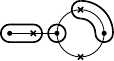}&$q_2^{(\text{I})}\negphantom{q_2^{(\text{I})}}\hphantom{Q_2^{(\text{I})}}$ \includegraphics[valign=c,padding = 0ex 0.5ex 0ex 0.5ex]{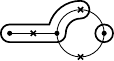}&$q_3^{(\text{I})}\negphantom{q_3^{(\text{I})}}\hphantom{Q_2^{(\text{I})}}$ \includegraphics[valign=c,padding = 0ex 0.5ex 0ex 0.5ex]{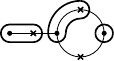}
    
     \\
     &
     $\tilde{q}_1^{(\text{I})}$ \includegraphics[valign=c,padding = 0ex 0.5ex 0ex 0.5ex]{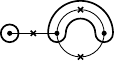}&$\tilde{q}_2^{(\text{I})}$ \includegraphics[valign=c,padding = 0ex 0.5ex 0ex 0.5ex]{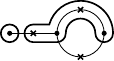}&$\tilde{q}_3^{(\text{I})} \negphantom{\tilde{q}_3^{(\text{I})}}\hphantom{Q_2^{(\text{I})}}$ \includegraphics[valign=c,padding = 0ex 0.5ex 0ex 0.5ex]{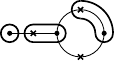}\\\hline&
     $g^{(\text{I})}\negphantom{g^{(\text{I})}}\hphantom{Q_2^{(\text{I})}}$ \includegraphics[valign=c,padding = 0ex 0.5ex 0ex 0.5ex]{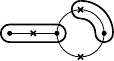}&$Z^{(\text{I})}\negphantom{Z^{(\text{I})}}\hphantom{Q_2^{(\text{I})}}$ \includegraphics[valign=c,padding = 0ex 0.5ex 0ex 0.5ex]{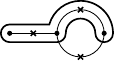}&$\tilde{g}^{(\text{I})}\negphantom{\tilde{g}^{(\text{I})}}\hphantom{Q_2^{(\text{I})}}$ \includegraphics[valign=c,padding =0ex 0.5ex 0ex 0.5ex]{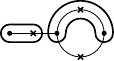}
\end{tabular}
\end{equation}
Like for the one-loop bubble, we have additional type II and III functions that are related to the type I functions. %The total number of these forms is forty which is four times the number of forms for the bubble, as predicted in \eqref{eq:Npancount}\hg{Added}.
The differential equations for each of these functions is easily derived from the kinematic flow rules.

\vskip 4pt
As a representative example, we consider the function $Q_1$. The tree for this function is
\begin{equation}
    Q_1:\raisebox{\height/4}{\includegraphics[valign=c]{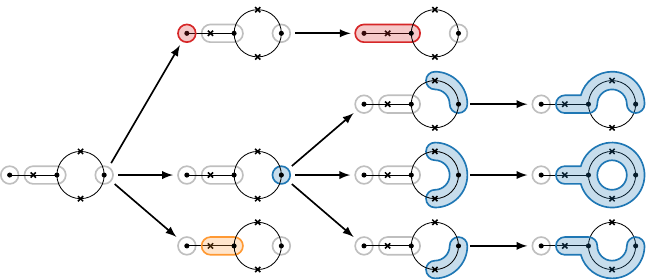}}
\end{equation}
In the top branch, we have the standard activation, growth and merger, while, in the bottom branch, we have only activation. In the middle branch, we have activation, then three types of growth and absorption. In principle, there could have been additional absorptions, but these lead to vanishing functions. The differential equation produced from this tree is
\begin{equation}
\begin{aligned}
    \ud Q_1&=\alpha_1\left(Q_1-q_1\right)\includegraphics[valign=c]{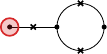}+\alpha_1\,q_1\,\includegraphics[valign=c]{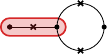} +\alpha_2\, Q_1\,\includegraphics[valign=c]{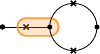}
    \\&+\alpha_3\left(
    Q_1-\tilde{q}_3^{(\text{I})}-\tilde{q}_3^{(\text{II})}-\tilde{q}_3^{(\text{III})}\right)\includegraphics[valign=c]{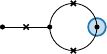}
    \\&+\alpha_3\hskip 2pt\left(\hskip 2pt\tilde{q}_3^{(\text{I})}\hskip 2pt +\hskip 2pt \tilde{q}_2^{(\text{I})} \hskip 2pt\right)\hskip 2pt\includegraphics[valign=c]{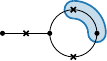}\hskip 1.5pt-\hskip 1.5pt\alpha_3\,\tilde{q}_2^{(\text{I})}\hskip 4pt\,\includegraphics[valign=c]{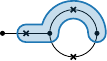}
    \\&+\alpha_3\hskip 1pt\left(\hskip 1pt\tilde{q}_3^{(\text{II})}\hskip 1pt +\hskip 1pt\tilde{q}_2^{(\text{II})}\hskip 1pt\right)\hskip 1pt\includegraphics[valign=c]{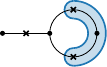}\hskip 1pt-\alpha_3\,\tilde{q}_2^{(\text{II})}\hskip 2pt\,\includegraphics[valign=c]{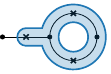}
    \\&+\alpha_3\left(\tilde{q}_3^{(\text{III})}+\tilde{q}_2^{(\text{III})}\right)\includegraphics[valign=c]{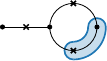} -\color{black}\alpha_3\,\tilde{q}_2^{(\text{III})}\,\includegraphics[valign=c]{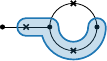}\, .
\end{aligned}
\end{equation}

\vskip 4pt
As a second example, we look at the function $Q_3^{({\rm I})}$. Its tree is
\beq
Q_3^{(\text{I})}:\includegraphics[valign=c]{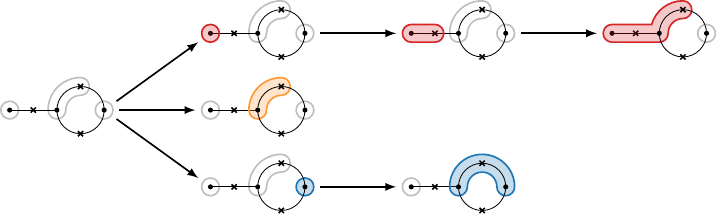}
\label{equ:Q3-tree}
\eeq
In the top branch, we see activation, growth and then absorption; in the middle branch, we have only activation; and, in the bottom branch, we have activation, growth and merger---we only have one growth/merger, instead of three, since the other two possibilities lead to vanishing functions.

\vskip 4pt
The differential equation associated to the tree (\ref{equ:Q3-tree}) is
\begin{equation}
\begin{aligned}
    \ud Q_3^{(\text{I})}&=\alpha_1\left(Q_3^{(\text{I})}-q_3^{(\text{I})}\right)\includegraphics[valign=c]{Frying_Pan_Forms_and_Letters/L1.pdf}+\alpha_1\left(q_3^{(\text{I})}+q_2^{(\text{I})}\right)\includegraphics[valign=c]{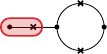}-\alpha_1q_2^{(\text{I})}\includegraphics[valign=c]{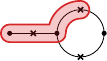}\\&\hskip 50.5pt+\alpha_2 Q_3^{(\text{I})}\includegraphics[valign=c]{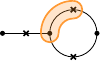}+\alpha_3\left(Q_3^{(\text{I})}-\tilde{q}_1^{(\text{I})}\right)\includegraphics[valign=c]{Frying_Pan_Forms_and_Letters/L4.pdf}+\alpha_3\tilde{q}_1^{(\text{I})}\includegraphics[valign=c]{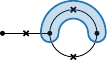}\, .
\end{aligned}
\end{equation}
We leave it as a straightforward exercise to the reader to derive the differential equations for the remaining source functions from the kinematic flow rules.

\subsubsection{Two-Loop Sunset}

Lastly, we present a two-loop example. The sunset graph has the following energy singularities:
\begin{equation}
\begin{aligned}
    \hat B_1&=\includegraphics[valign=c]{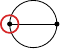}=X_1+Y_1+Y_2+Y_3\,,& \hat B_2&=\includegraphics[valign=c]{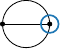}=X_2+Y_1+Y_2+Y_3\,,&\\
    \hat B_3&=\includegraphics[valign=c]{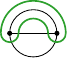}=X_1+X_2+2Y_2+2Y_3\,,& \hat B_4&=\includegraphics[valign=c]{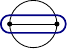}=X_1+X_2+2Y_1+2Y_3\,,\\ \hat B_5&=\includegraphics[valign=c]{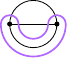}=X_1+X_2+2Y_1+2Y_2\,,&
    \hat B_6&=\includegraphics[valign=c]{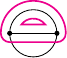}=X_1+X_2+2Y_3\,,&\\ \hat B_7&=\includegraphics[valign=c]{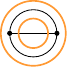}=X_1+X_2+2Y_2\,,&\hat B_8&=\includegraphics[valign=c]{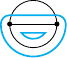}=X_1+X_2+2Y_1\,,\\ \hat B_9&=\includegraphics[valign=c]{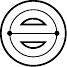}=X_1+X_2\,. 
\end{aligned}
\end{equation}
The flat-space wavefunction coefficient can then be constructed from the sum of all compatible complete tubings, as defined in~\eqref{eq:FlatWFC}. Each such tubing consists of four nested tubes and so displaying them graphically is challenging.

\vskip 4pt
For this system, we have the following basis functions:
\begin{equation}\label{tab:sunforms}
    \begin{tabular}{p{1.69cm}p{1.69cm}p{1.69cm}p{1.69cm}p{1.69cm}p{1.69cm}p{1.69cm}}
    $F_{00}$\,\includegraphics[valign=c,padding = 0ex 0.5ex 0ex 0.5ex]{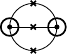}\\\hline
    $F_{10}$\,\includegraphics[valign=c,padding = 0ex 0ex 0ex 0.5ex]{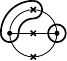}&
    $F_{20}$\,\includegraphics[valign=c,padding = 0ex 0ex 0ex 0.5ex]{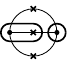}&
    $F_{30}$\,\includegraphics[valign=c,padding = 0ex 0ex 0ex 0.5ex]{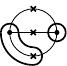}&
    $F_{40}$\,\includegraphics[valign=c,padding = 0ex 0ex 0ex 0.5ex]{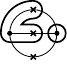}&
    $F_{50}$\,\includegraphics[valign=c,padding = 0ex 0ex 0ex 0.5ex]{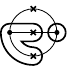}&
    $F_{60}$\,\includegraphics[valign=c,padding = 0ex 0ex 0ex 0.5ex]{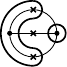}&
    $F_{70}$\,\includegraphics[valign=c,padding = 0ex 0ex 0ex 0.5ex]{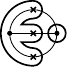}\\
    $F_{01}$\,\scalebox{-1}[1]{\includegraphics[valign=c,padding = 0ex 0.5ex 0ex 0ex]{Sunset_Forms/F10.pdf}}&
    $F_{02}$\,\scalebox{-1}[1]{\includegraphics[valign=c,padding = 0ex 0.5ex 0ex 0ex]{Sunset_Forms/F20.pdf}}&
    $F_{03}$\,\scalebox{-1}[1]{\includegraphics[valign=c,padding = 0ex 0.5ex 0ex 0ex]{Sunset_Forms/F30.pdf}}&
    $F_{04}$\,\scalebox{-1}[1]{\includegraphics[valign=c,padding = 0ex 0.5ex 0ex 0ex]{Sunset_Forms/F40.pdf}}&
    $F_{05}$\,\scalebox{-1}[1]{\includegraphics[valign=c,padding = 0ex 0.5ex 0ex 0ex]{Sunset_Forms/F50.pdf}}&
    $F_{06}$\,\scalebox{-1}[1]{\includegraphics[valign=c,padding = 0ex 0.5ex 0ex 0ex]{Sunset_Forms/F60.pdf}}&
    $F_{07}$\,\scalebox{-1}[1]{\includegraphics[valign=c,padding = 0ex 0.5ex 0ex 0ex]{Sunset_Forms/F70.pdf}}\\\hline
    $F_{11}$\,\includegraphics[valign=c,padding = 0ex 0.5ex 0ex 0.5ex]{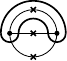}&
    $F_{22}$\,\includegraphics[valign=c,padding = 0ex 0.5ex 0ex 0.5ex]{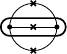}&
    $F_{33}$\,\includegraphics[valign=c,padding = 0ex 0.5ex 0ex 0.5ex]{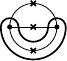}&
    $F_{44}$\,\includegraphics[valign=c,padding = 0ex 0.5ex 0ex 0.5ex]{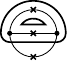}&
    $F_{55}$\,\includegraphics[valign=c,padding = 0ex 0.5ex 0ex 0.5ex]{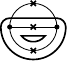}&
    $F_{66}$\,\includegraphics[valign=c,padding = 0ex 0.5ex 0ex 0.5ex]{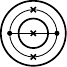}&
    $F_{77}$\,\includegraphics[valign=c,padding = 0ex 0.5ex 0ex 0.5ex]{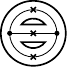}
\end{tabular}
\end{equation}
where $F_{00}=\psi$. The vanishing functions come in 6 different types:
\begin{equation}
    \includegraphics[valign=c]{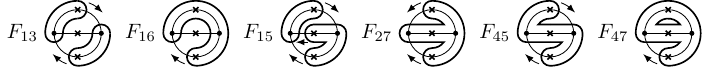}
\end{equation}
In total, there are therefore 22 basis functions with 1, 14, 7 functions at levels 0, 1, 2.
Just as for the one-loop bubble, these function split into sets of three, $F_{i0},\, F_{0i}$ and $F_{ii}$, which act independently of each other. The differential equations for all basis functions is easily derived from the kinematic flow:
\begin{align}\nonumber
    \ud\psi&=\alpha_1\left[\psi\,  \ud\log \hat B_1+\sum_{i}F_{i0}\left(\ud\log X_1^{a_ib_ic_i}-\ud\log \hat B_1\right)\right] \\\label{eq:sun1}
    &+ \alpha_2\left[\psi\, \ud\log \hat B_2+\sum_{i}F_{0i}\left(\ud\log X_2^{a_ib_ic_i}-\ud\log \hat B_2\right)\right] , \\\label{eq:sun2}
    \ud F_{i0}&=\alpha_1F_{i0}\,\ud\log X_1^{a_ib_ic_i}+\alpha_2\left[(F_{i0}-F_{ii})\ud\log \hat B_2+F_{ii}\ud\log \hat B_{2+i}\right] , \\\label{eq:sun3}
    \ud F_{0i}&=\alpha_1\left[(F_{0i}-F_{ii})\, \ud\log \hat B_1+F_{ii}\ud\log \hat B_{2+i}\right]+\alpha_2 F_{0i}\,\ud\log X_{2}^{a_ib_ic_i}\,,\\\label{eq:sun4}
    \ud F_{ii}&=(\alpha_1+\alpha_2)F_{ii}\,\ud\log \hat B_{2+i}\,,
\end{align}
where
\begin{align}
X_i^{a_ib_ic_i}&=X_i+a_iY_1+b_iY_2+c_iY_3\,,
\end{align}
with all $a_i,\, b_i,\, c_i=1$ except for 
\beq
\begin{aligned}
    a_1&=-1\,, \qquad  a_4=b_4=-1\,, \qquad  a_7=b_7=c_7=-1\,. \\
    b_2&=-1\,, \qquad b_5=c_5=-1\,, \\
    c_3&=-1\,, \qquad   a_6=c_6=-1\,.
\end{aligned}
\eeq
These equations are easily generalized
 for arbitrary $n$-loop two-site graphs.

%\newpage
\section{Conclusions and Outlook}
\label{sec:Conclusions}

Conformally coupled scalars in a power-law FRW cosmology are an interesting toy model. The simplicity of its mode functions allows a large class of correlators to be computed explicitly and the resulting mathematical data can then be explored to find hidden structures.
In~\cite{Arkani-Hamed:2023kig,Arkani-Hamed:2023bsv}, it was found that all tree-level correlators---of arbitrary multiplicity---obey simple and universal graphical rules describing a flow in kinematic space. In this note, we showed that the same rules also predict the differential equations for loop integrands in this theory, if we account for the fact that certain basis integrals can vanish. We demonstrated this for %selected 
one- and two-loop examples.

\vskip 10pt
There are a number of open problems in the exploration of the ``kinematic flow" that remain to be addressed: 
\begin{itemize}
\item First of all, although we have shown that the flow rules predict the differential equations for loop integrands, performing the actual loop ``integrals" remains an important challenge (see~\cite{Benincasa:2024lxe, Benincasa:2024ptf}
for recent progress). It would be fascinating if something like the kinematic flow also applied after loop integration.
\item Next, the patterns that we discovered, so far, have only been for conformally coupled scalars in a power-law cosmology.
Fields with general masses pose a challenge, because their mode functions are Hankel functions. Using an integral representation of these Hankel functions, however, lets the wavefunction coefficients again be written as twisted integrals that satisfy similar differential equations. It would be interesting to see if these equations obey a generalized version of the kinematic flow.
\item Finally, and most importantly, we don't actually know ``why" the kinematic flow works. So far, the rules for deriving the differential equations were discovered purely ``experimentally". It would be interesting to find a deeper geometrical structure from which the properties of the flow can actually be derived. 
\end{itemize}
We hope to return to some of these questions in future work.

 \vspace{0.2cm}
 \paragraph{Acknowledgments} DB and HL thank Nima Arkani-Hamed,  Aaron Hillman, Austin Joyce and Guilherme Pimentel for collaboration on the ``kinematic flow" which inspired the present work. DB and HG are grateful to Sharan Srinivasan for initial collaboration.
 
 \vskip 4pt
The research of DB is funded by the European Union (ERC,  \raisebox{-2pt}{\includegraphics[height=0.9\baselineskip]{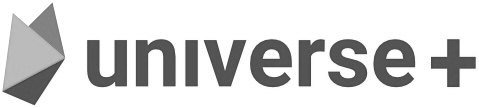}}, 101118787). Views and opinions expressed are, however, those of the author(s) only and do not necessarily reflect those of the European Union or the European Research Council Executive Agency. Neither the European Union nor the granting authority can be held responsible for them. DB is further supported by a Yushan Professorship at National Taiwan University funded by the Ministry of Education (MOE) NTU-112V2004-1. DB thanks the Max-Planck-Institute for Physics (MPP) in Garching for its hospitality while this work was being completed. He is grateful to the Alexander von Humboldt-Stiftung and the Carl Friedrich von Siemens-Stiftung for supporting his visits to the~MPP. HG is supported by a Postdoctoral Fellowship at National Taiwan University funded by the National Science and Technology Council (NSTC) 113-2811-M-002-073. HG was also supported by a Postdoctoral Fellowship at National Taiwan University funded by the Ministry of
Education (MOE) NTU-112L4000-1. HL was supported by the Kavli Institute for Cosmological Physics at the University of Chicago.

%\newpage
%%%%%%%%%%%%%%%%%%%%%%%%%
\vspace{0.5cm}
\phantomsection
%\enlargethispage{\baselineskip}
%\addtocontents{toc}{\protect\enlargethispage{\baselineskip}}
\addcontentsline{toc}{section}{References}
\bibliographystyle{utphys}
{\linespread{1.075}
	\bibliography{Loops-Refs}
}
\end{document}